\documentclass[12pt]{article}
\usepackage[margin=1in]{geometry}
\usepackage{textcomp}
\usepackage{iftex}
\ifPDFTeX
 \usepackage[utf8]{inputenc}
 \usepackage[T1]{fontenc}
\else
 \usepackage{fontspec}
\fi
\usepackage{float}
\usepackage{booktabs}
\usepackage{longtable}
\usepackage{tabularx}
\usepackage{array}
\usepackage{placeins}
\usepackage{graphicx}
\usepackage{caption}
\usepackage[caption=false]{subfig}
\usepackage[longnamesfirst,sort]{natbib}
\bibpunct[, ]{(}{)}{;}{a}{,}{,}

\usepackage{xcolor}

\usepackage{hyperref}
\usepackage{pdflscape}
\usepackage{fvextra}
\usepackage{array}
\newenvironment{keywords}{\vspace{0.5em}\noindent\textbf{Keywords: }}{\par\vspace{1em}}

\begin{document}

\begin{center}

{\Large \textbf{MedEasy: Designing AI Standardized Patients for Clinical Consultation Training}\par}

\vspace{1.0em}

{\normalsize
\mbox{Zhiqi Gao\textsuperscript{1,\textdagger}},
\mbox{Huarui Luo\textsuperscript{2,\textdagger}},
\mbox{Guo Zhu\textsuperscript{3,\textdagger}},
\mbox{Bingquan Zhang\textsuperscript{1}}\\[0.35em]
\mbox{Dongyijie Primo Pan\textsuperscript{4}},
\mbox{Yizhan Feng\textsuperscript{5}},
\mbox{Jiahuan Pei\textsuperscript{6}},
\mbox{Jie Li\textsuperscript{7,*}},
\mbox{Benyou Wang\textsuperscript{1,*}}
\par}

\vspace{1.0em}

{\footnotesize
\textsuperscript{1}School of Data Science, The Chinese University of Hong Kong, Shenzhen, Shenzhen, China\\
\textsuperscript{2}School of Medicine, Nankai University, Tianjin, China\\
\textsuperscript{3}School of Artificial Intelligence, The Chinese University of Hong Kong, Shenzhen, Shenzhen, China\\
\textsuperscript{4}Information Hub, The Hong Kong University of Science and Technology (Guangzhou), Guangzhou, China\\
\textsuperscript{5}Department of Mathematics, The Chinese University of Hong Kong, Shenzhen, Shenzhen, China\\
\textsuperscript{6}Department of Artificial Intelligence, Vrije Universiteit Amsterdam, Amsterdam, The Netherlands\\
\textsuperscript{7}MIT Media Lab, Massachusetts Institute of Technology, Cambridge, MA, USA
\par}

\vspace{0.8em}
{\footnotesize
\textsuperscript{\textdagger}These authors contributed equally to this work.\\
\textsuperscript{*}Corresponding authors.\\[0.4em]
Author emails: Zhiqi Gao, \href{mailto:gaozhiqi@mail.nankai.edu.cn}{gaozhiqi@mail.nankai.edu.cn};
Huarui Luo, \href{mailto:2120251953@mail.nankai.edu.cn}{2120251953@mail.nankai.edu.cn};
Guo Zhu, \href{mailto:guozhu@link.cuhk.edu.cn}{guozhu@link.cuhk.edu.cn};
Bingquan Zhang, \href{mailto:zhangbingquan@cuhk.edu.cn}{zhangbingquan@cuhk.edu.cn};
Dongyijie Primo Pan, \href{mailto:dpan750@connect.hkust-gz.edu.cn}{dpan750@connect.hkust-gz.edu.cn};
Yizhan Feng, \href{mailto:yizhanfeng@link.cuhk.edu.cn}{yizhanfeng@link.cuhk.edu.cn};
Jiahuan Pei, \href{mailto:j.pei2@vu.nl}{j.pei2@vu.nl};
Jie Li, \href{mailto:jieli8@mit.edu}{jieli8@mit.edu};
Benyou Wang, \href{mailto:wangbenyou@cuhk.edu.cn}{wangbenyou@cuhk.edu.cn}.
\par}

\vspace{0.8em}

{\footnotesize Word count: 15734\par}

\end{center}

\vspace{1.2em}
\vspace{1.2em}

\begin{abstract}
AI standardized patients are becoming a setting for professional training in clinical consultation. This paper presents MedEasy, a multi-agent system that organizes virtual-patient practice through patient dialogue, clinical actions, decision submission, documentation, and feedback. We first conducted a formative study with 12 clinical-year medical students through interviews and three co-design workshops. The findings informed a staged workflow, structured case records, action-contingent findings, and trajectory-based review. We then conducted an evaluative user study with a separate cohort of 12 clinical-year medical students, with each participant completing two counterbalanced cases. Learners interpreted MedEasy as a connected consultation environment. They used patient responses, examination findings, available actions, and feedback together to judge whether the represented case remained coherent. They valued repeatable practice and recorded review, while questioning missing actions and feedback criteria. The paper contributes design implications for AI-supported professional training systems that use case-specific standards to connect situated practice.
\end{abstract}


\section{Introduction}

Recent advances in large language models (LLMs) are changing the forms of clinical consultation practice available to medical learners. Clinical consultation requires learners to elicit patient concerns, interpret incomplete information, choose examinations, explain decisions, and communicate in understandable ways \citep{epstein2007patient,makoul2001essential,rider2006communication,silverman2016skills}. These abilities develop through repeated interaction in settings that approximate clinical encounters \citep{barrows1993overview,cleland2009use,nestel2014simulated}. Human standardized patients (SPs) have long provided such practice without placing real patients at unnecessary risk \citep{barrows1993overview,cleland2009use,nestel2014simulated,rutherford2024use}. Their delivery depends on trained actors, scheduling, case preparation, and educator time \citep{elendu2024impact,flanagan2023standardized,issenberg2005features}. Virtual patients extended access through scripted dialogue and predefined case progression \citep{cook2009virtual,cook2010computerized,kononowicz2015virtual}. Recent AI standardized patients further extend this work through generative dialogue, feedback, and embodied or multimodal interaction \citep{borg2024creating,cross2025chatgptsp,guetterman2019mpathicvr,Hicke_2026,louie2026can,sehgal2025pal,wang2025generative}. Existing evaluations have examined conversational fluency, persona consistency, emotional responsiveness, and perceived realism \citep{borg2024creating,brown2020language,holderried2024generative,tu2024towards,zhu2025vaps}.

Clinical consultation practice, however, is not contained in a single patient response. A learner moves between patient accounts, examination findings, test results, diagnostic possibilities, management decisions, and later review \citep{amershi2019guidelines,eva2005every,makoul2001essential,silverman2016skills}. In an AI-SP system, these parts may be handled by different components. A generated response may disclose information too early, omit a requested detail, conflict with authored case information, or agree too readily with the learner's suggestion \citep{bender2021dangers,ferrario2024role,ghassemi2021false}. When this happens, the learner may need to judge whether the issue comes from the case design, the system's interpretation of the question, response generation, the interface, or their own clinical reasoning \citep{dietvorst2015algorithm,parasuraman1997humans}. The consultation represented by an AI-SP is therefore assembled across dialogue, case evidence, clinical actions, and evaluation.

This paper examines how medical learners interpret such assembled consultation practice. Existing research has provided evidence about AI-SP realism, usability, acceptance, and educational potential. System-oriented work has also introduced intent-conditioned disclosure, modular patient simulation, and trajectory-level feedback. Less is known about how learners make sense of the relationships among generated patient responses, authored case information, available clinical actions, and post-session evaluation. Learners also bring prior experiences with simulation and AI-mediated learning into the session. These experiences can shape what they expect an AI-SP to do, what they treat as a meaningful clinical action, and how they read automated feedback. We therefore ask:

\begin{itemize}
 \item \textbf{RQ1:} How do medical learners interpret MedEasy in relation to their prior experiences with human standardized patients, virtual patients, and AI-mediated learning tools?
 \item \textbf{RQ2:} How do medical learners use and evaluate MedEasy as a consultation training environment?
\end{itemize}

To examine these questions, we studied MedEasy, an implemented AI-SP system that separates consultation practice into dialogue, authored case evidence, learner-selected clinical actions, and post-session evaluation. During history taking, an Auxiliary Agent maps each learner question to one or more predefined clinical intents, and a Patient Agent generates a response using the corresponding authored case information. Physical-examination and auxiliary-test findings are returned after explicit learner actions. Learners then complete diagnosis, patient education, treatment planning, and structured medical-record writing. After the session, an Evaluation Agent compares the recorded dialogue, elicited facts, clinical actions, and submitted decisions with case-specific criteria to produce structured feedback. In this paper, MedEasy serves as the empirical site for studying how learners interpret a consultation distributed across these components.

We conducted a multi-phase qualitative design study with clinical-year medical students. Twelve learners first participated in formative interviews and three co-design workshops. A separate group of 12 learners later completed two counterbalanced cases with the implemented system and took part in semi-structured interviews. We analyzed the formative materials as design-stage evidence and the later interviews as an independent evaluative user-study dataset. Prior simulation and AI-mediated learning experiences were discussed in interviews as participants' points of reference; they were not treated as standardized comparison conditions.

The present paper builds on two prior publications from the same research program. \citet{gao2026talks} reported the formative interviews and co-design activities that first identified learner requirements for AI-SP consultation practice. \citet{zhang2026easymed} described the implemented MedEasy system architecture. This paper brings these strands together through a new qualitative design account. We reanalyze and organize the formative materials as design-stage evidence, trace how the resulting requirements were translated into the implemented MedEasy workflow, and report an evaluative user study with a separate participant cohort. 

This paper makes three contributions. First, it provides a design-stage account of how formative interviews and co-design workshops were synthesized into requirements for AI-SP consultation practice beyond patient dialogue. Second, it traces how these requirements were implemented in MedEasy through a staged workflow, structured case representation, clinical-action interfaces, and post-session feedback. Third, it reports an evaluative user study showing how a new cohort of medical learners interpreted MedEasy as a consultation training environment where dialogue, clinical action, and automated review were connected in use.

\section{Background and Related Work}

We review three areas of work that frame our study: standardized and AI-simulated patients, trust and feedback in AI-mediated professional training, and clinical consultation practice beyond patient dialogue.

\subsection{Standardized Patients and AI-Driven Patient Simulation}

Medical education has long used simulated encounters to support clinical communication and reasoning. Human standardized patients (SPs) allow learners to practice clinical encounters without exposing real patients to unnecessary risk \citep{barrows1993overview,cleland2009use,nestel2014simulated}. Their bodily presence and responses to learner behavior are especially important when an encounter involves interpersonal pressure, non-verbal communication, or changes in patient attitude \citep{cleland2009use,nestel2014simulated}. SP programs depend on actor training, case preparation, scheduling, supervision, and consistent performance \citep{elendu2024impact,flanagan2023standardized}. The educational value of SP feedback also depends on how comments are delivered and connected to the learner's performance \citep{cleland2009use,schlegel2012feedback}. These practical demands often concentrate SP encounters within scheduled teaching and assessment, limiting opportunities for repeated and self-directed practice \citep{cleland2009use,flanagan2023standardized,issenberg2005features}.

Virtual patients provide a more accessible and repeatable way to engage with clinical cases and have become an established component of medical education \citep{cook2009virtual,cook2010computerized,berman2016role}. Earlier systems commonly relied on scripted dialogue, branching scenarios, or predefined case pathways \citep{huang2007virtual}. Their educational value depends not only on how the patient is represented, but also on how the scenario, learner action, and feedback are organized within the training activity \citep{lee2020virtualpatient}.

Large language models have expanded virtual-patient interaction through open-ended dialogue and context-sensitive responses \citep{brown2020language,holderried2024generative,zhu2025vaps}. Recent AI-SP systems have applied these capabilities to clinical rehearsal and formative feedback \citep{borg2024creating,du2024llms,grevisse2024raspatient,Hicke_2026,sehgal2025pal,wang2025generative}. Some systems combine patient dialogue with post-session summaries and clinically oriented learning materials \citep{li2024leveraging}. Other studies have examined whether LLM-based patients can represent emotionally demanding or challenging clinical interactions \citep{bodonhelyi2025modeling}.

These developments extend the range of interactions available in virtual-patient training, while retaining several limitations. AI-simulated patients may provide limited affective cues, respond too cooperatively, introduce information inconsistent with the authored case, or vary in persona and clinical plausibility across turns \citep{borg2024creating,ferrario2024role,tu2024towards}. Accordingly, AI-SP practice cannot be assessed only through conversational fluency or perceived realism. It also depends on how patient information is grounded, how the case develops, and how learner action is interpreted within the training activity.

\subsection{Trust, Reliability, and Feedback in AI-Mediated Training}

Trust in automation and AI is associated with users' assessments of reliability, predictability, competence, and transparency under uncertainty \citep{lee2004trust,mayer1995integrative,rousseau1998not}. Research distinguishes general attitudes toward a technology from assessments formed through experience with its actual capabilities \citep{hoff2015trust,lee2004trust,zhang2020effect}. These assessments change as users observe system outputs, encounter errors, and revise their expectations \citep{buccinca2021trust,de2020towards}. Excessive confidence may lead users to accept unreliable outputs, while rejection following an error may lead them to disregard support that remains useful in other situations \citep{dietvorst2015algorithm,parasuraman1997humans}. HCI research has therefore examined interfaces that communicate system status and allow users to question, correct, or recover from uncertain behavior \citep{amershi2019guidelines,bansal2019beyond,bansal2021does}.

In professional training, learners assess both the simulated activity and the system's interpretation of their performance. They may consider whether a patient response is clinically plausible, whether the available evidence is internally consistent, and whether the system has correctly recorded an action. The same training environment may later evaluate the learner using a predefined case representation or rubric. Trust in the simulated encounter and trust in its evaluation are connected, but they may develop differently. A plausible patient interaction does not establish that the subsequent feedback is complete or clinically appropriate.

Educational feedback compares observed performance with a reference standard and informs subsequent learning \citep{vanderidder2008feedback,ramani2019feedback}. In AI-mediated clinical training, reference standards may specify expected questions, examinations, diagnostic conclusions, or treatment decisions. Automated evaluation can identify recorded actions and omissions, but its judgments depend on the content and scope of the supplied standard. Criteria designed for one learner stage, specialty context, or teaching objective may not transfer directly to another setting.

Prior research on AI in education has similarly shown that automated judgments are shaped by the data, pedagogical assumptions, and institutional contexts through which they are produced \citep{holstein2019designing,ouyang2021artificial,zawacki2019systematic}. For AI-SP training, the credibility of feedback therefore depends on more than the fluency of its explanation. Learners and educators may also need to understand which action prompted a comment, which case criterion was applied, and whether that criterion fits the clinical context in which the decision was made.

\subsection{Clinical Consultation Practice beyond Patient Dialogue}

AI-SP research has often examined natural dialogue, emotional expression, and persona consistency as conditions for engagement and perceived realism \citep{nass2000machines,holderried2024generative,zhu2025vaps}. Clinical consultation practice also requires learners to connect patient accounts with clinical evidence, decisions, and explanation \citep{makoul2001essential,silverman2016skills}. These activities unfold as new evidence becomes available and provisional interpretations are revised \citep{eva2005every}. Patient dialogue is one part of a broader practice task in which information and decisions develop across the encounter.

In an AI-mediated training environment, the design determines what clinical findings become available, what learners can do, and how their work is recorded for later review. Patient-reported information may be generated through dialogue, while examination and test findings may come from predefined case materials. Diagnostic and management decisions may then be evaluated against a separate reference standard. Learners encounter these elements as parts of the same case, even when they originate from different representations or computational processes.

Human--AI interaction research has emphasized visible system status, understandable capabilities, and recovery from uncertain system behavior \citep{amershi2019guidelines,bansal2021does,buccinca2021trust,kocielnik2019will,liao2022designing}. In AI-SP practice, these concerns extend across the consultation. Learners may need to understand how their questions, actions, and later feedback are connected within the represented case.

Recent medical simulation systems have begun to represent consultation activities beyond dialogue through structured case progression, clinical actions, and automated feedback \citep{ouyang2026casemaster,steenstra2025simpatient}. Multi-agent clinical simulations have examined coordination among specialized components within patient simulation \citep{almansoori2025selfevolvingmultiagentsimulationsrealistic}. Related work has also described expert principles and constraints for constructing LLM-simulated patients \citep{park2024roleplay}. Research on intelligent tutoring and simulation further shows that learners need understandable task progression when they make errors or follow unexpected paths \citep{rzepka2018user}.

Together, this work shows that AI-SP systems are beginning to represent consultation as more than patient dialogue. However, existing studies still provide limited accounts of how learners interpret the consultation they encounter when different parts of the activity are handled by different system components. In particular, we know less about how learners make sense of inconsistencies, system boundaries, and evaluation criteria in such settings.

\section{Formative Study and Design Translation}
\label{sec:formative}

\begin{figure}[!htb]
\centering
\includegraphics[width=\linewidth]{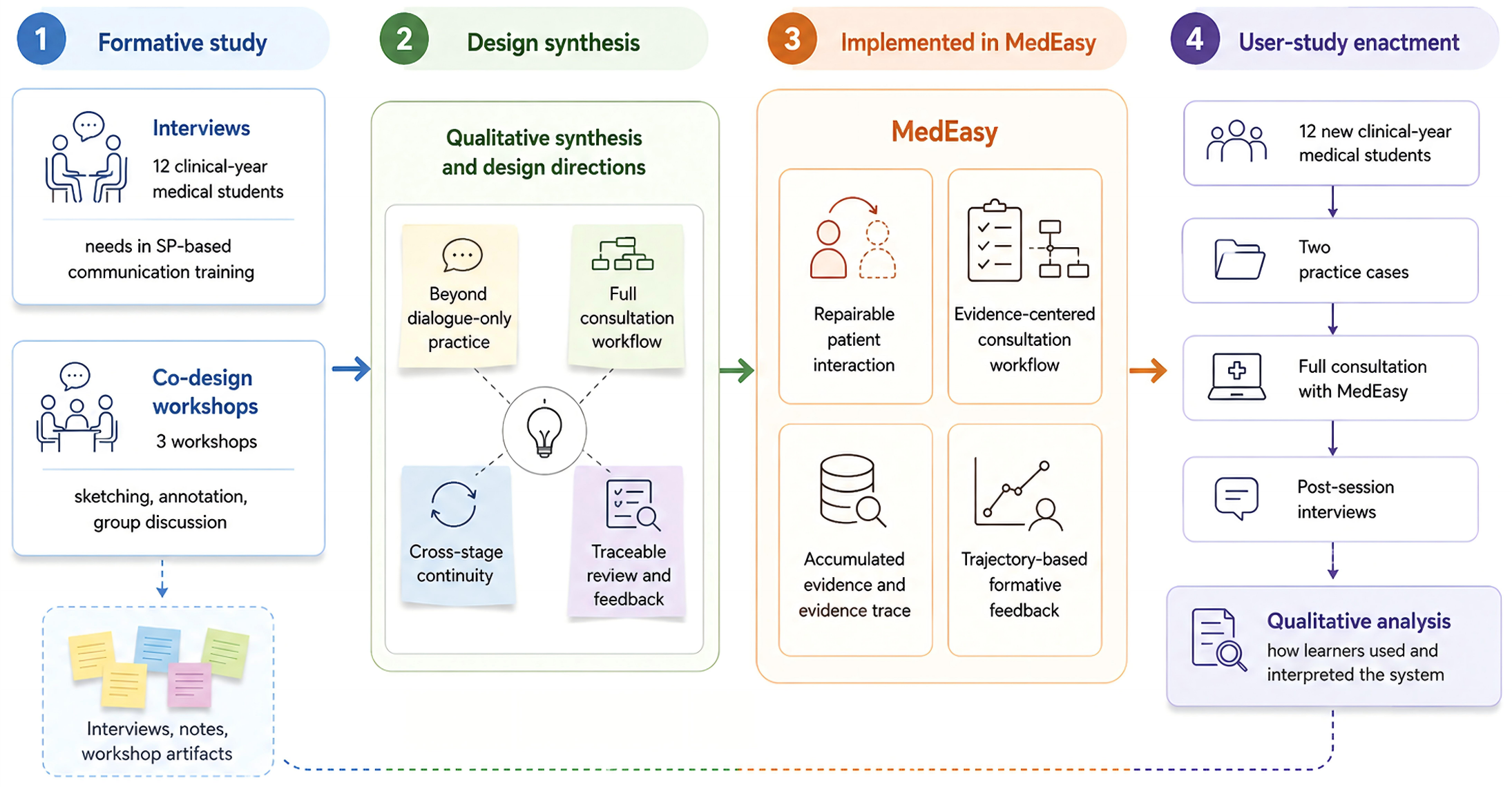}
\caption{Overview of the study process and design translation. The formative study combined interviews and co-design workshops with 12 clinical-year medical students. The resulting qualitative synthesis informed four design directions, which were implemented in MedEasy as repairable patient interaction, an evidence-centered consultation workflow, accumulated evidence with evidence trace, and trajectory-based formative feedback. An evaluative user study with a separate cohort examined how learners used and interpreted the implemented system.}
\label{fig:study-process}
\end{figure}

We conducted formative interviews and co-design workshops to understand learners' prior simulation experiences, identify limitations in early AI-SP practice, and inform MedEasy's consultation workflow, interaction design, and feedback presentation \citep{blandford2016qualitative,creswell2016qualitative,sanders2008cocreation,spinuzzi2005pd}. These activities took place between December 2025 and January 2026. All interviews and workshop discussions were conducted in Mandarin Chinese, audio-recorded, transcribed, and de-identified. Coding was performed on the original Chinese-language transcripts. Quotations selected for publication were translated into English during manuscript preparation and checked against the original transcripts. We identify formative participants as F1--F12 and workshop sessions as WS1--WS3. Figure~\ref{fig:study-process} summarizes how the formative study, design synthesis, MedEasy implementation, and evaluative user study were connected.

\subsection{Participants and Procedure}

\textbf{Participants.}
We recruited 12 medical students in Years 4--6 from two medical schools. All had prior experience with human SP encounters in clinical communication, physical examination, or OSCE-related training. Recruitment took place through institutional mailing lists and snowball sampling. Participation was voluntary, unrelated to academic assessment, and compensated. Participant characteristics are reported in Table~\ref{tab:participant-information}(a) in Appendix A.

\textbf{Interviews.}
Each participant completed an individual semi-structured interview by video conference. Interviews lasted approximately 45 minutes and were conducted in Mandarin Chinese. Participants first described previous human SP encounters, including how they gathered information, performed examinations, responded to patient behavior, and received feedback.

Participants then interacted with an early version of MedEasy that supported patient questioning and provided basic case, examination, and test information through predefined functions. After using the prototype, they discussed the patient dialogue, consistency of case information, available clinical actions, and expected forms of training feedback. The interviews concluded by examining which activities AI-SPs could support and which continued to require human SPs, educators, physical models, or real patients. The interview guide is included in Appendix D.

\textbf{Workshops.}
The 12 formative participants were divided into three groups of four, with each group attending one 75-minute in-person workshop conducted in Mandarin Chinese. Participants first produced an individual sketch of an AI-SP system for consultation training, considering the sequence of the encounter, forms of interaction, patient representation, and feedback. The design sheets were then circulated across two annotation rounds, allowing participants to comment on, extend, or question one another's proposals before a group discussion of the resulting designs.

\begin{figure}[!htb]
\centering
\includegraphics[width=\linewidth]{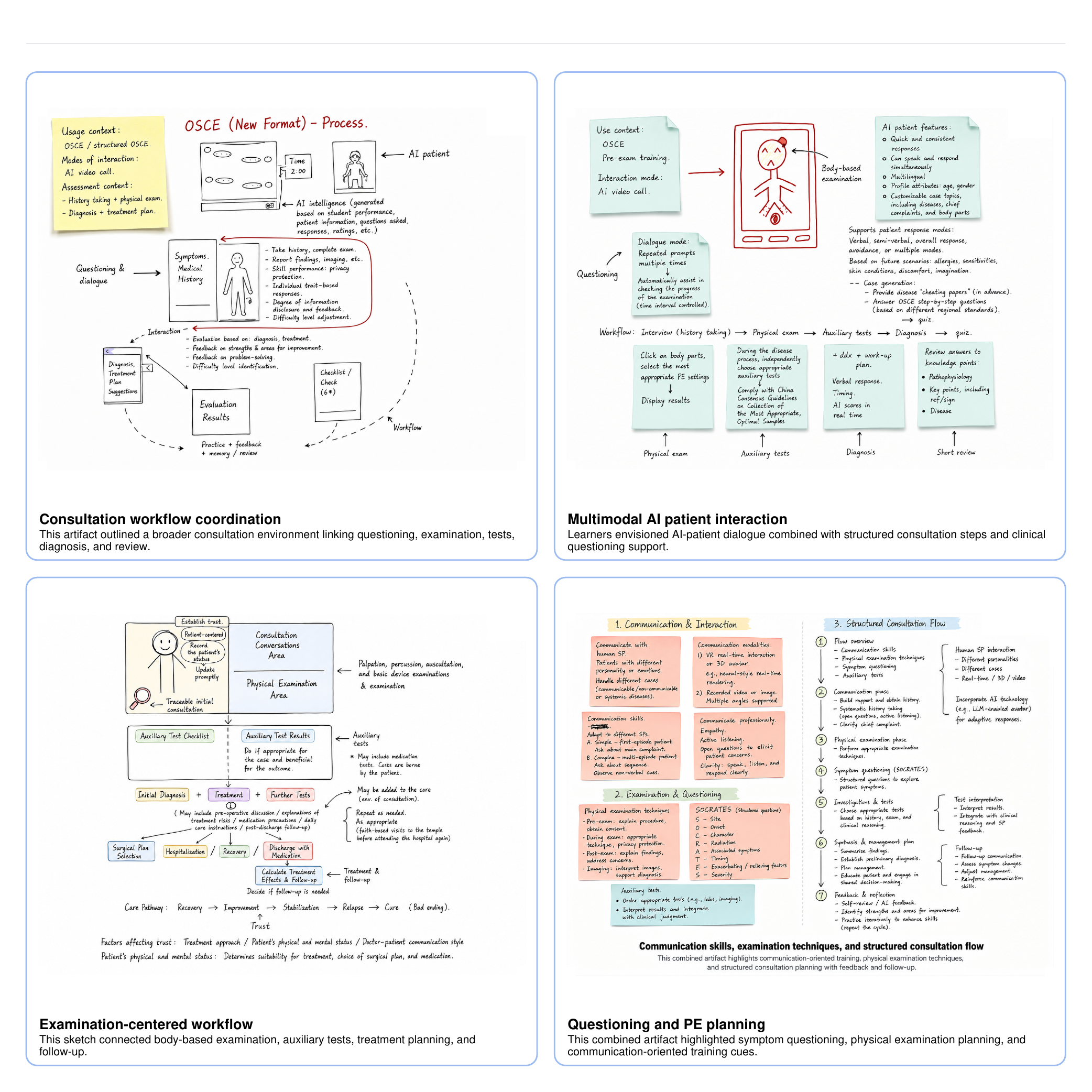}
\caption{Representative co-design artifacts from the formative workshops. Participants sketched AI-SP consultation environments connecting patient questioning, physical examination, auxiliary tests, diagnosis, treatment planning, and review. The original Chinese artifacts were translated into English and analyzed with participants' spoken explanations and annotations.}
\label{fig:codesign-artifacts}
\end{figure}

We collected the sketches and annotations, photographed the workshop materials, audio-recorded and transcribed the discussions, and retained facilitator notes. Visual artifacts were analyzed together with participants' spoken explanations. Figure~\ref{fig:codesign-artifacts} shows representative workshop artifacts that informed the later design synthesis.

\subsection{Formative Data Analysis}\label{sec:formative_analysis}
We analyzed the formative interviews and co-design workshop materials as design-stage evidence using a team-based deductive–inductive hybrid thematic analysis \citep{fereday2006demonstrating,xu2020applying}. The purpose of the analysis was to identify design-relevant user needs and concerns that could inform the next prototype iteration.

Two researchers developed an initial code manual as a sensitizing structure for extracting design-relevant evidence, rather than as a fixed scheme for determining the final requirements. The initial codes were derived from the interview guide and workshop tasks, covering areas such as consultation flow, patient dialogue, feedback expectations, and prior simulation experience. The researchers then read the formative interview transcripts and workshop materials to refine this structure and added recurring design issues not fully captured by the initial categories, such as unclear learner roles, missing action flexibility, and the need to connect dialogue with examination. 

For each code, the researchers defined inclusion and exclusion criteria to clarify which participant statements, workshop annotations, or spoken explanations should be treated as evidence for a potential design requirement. For example, comments about what feedback should provide, when it should appear, or how it should support learning were included under feedback requirements, whereas general remarks such as ``feedback is important'' were excluded unless they specified an actionable design need.

During coding, the researchers identified design-relevant meaning units: coherent segments of data expressing a user need, expectation, breakdown, workflow constraint, or suggestion. These meaning units were converted into statement cards, each containing a participant or workshop identifier, a supporting quotation or workshop note, a one-sentence analytic summary, the associated code, and a possible design implication. This process produced 235 statement cards: 169 from the formative interviews and 66 from the workshops.

During the inductive synthesis stage, the researchers compared and grouped the statement cards through affinity diagramming \citep{lucero2015affinity}. The initial code manual supported the first cycle of coding but did not determine the final thematic structure. The researchers reviewed emerging clusters against their linked quotations, discussed overlapping categories, and revised the groupings until each cluster captured a coherent design requirement. Through this process, four design-oriented themes emerged.

Because our goal was to develop design implications rather than to measure the distribution of predefined categories, we used consensus-based coding instead of calculating inter-rater reliability. Disagreements were treated as opportunities to refine code definitions and clarify design interpretations.



\subsection{Formative Study Findings}

The formative analysis produced four design-oriented themes. They concerned human SPs and real patients as reference practices, the scope of AI-SP consultation, the organization of activities across an encounter, and post-session review. These findings informed the design requirements for MedEasy.

\subsubsection{Human SPs and Real Patients as Reference Practices}

Participants did not evaluate AI-SPs only by asking whether a simulated patient could answer questions. They compared AI-mediated practice with human SP and real-patient encounters that involved bodily response, interpersonal pressure, patient attitude, access to practice, and feedback after the encounter. Human SPs therefore served as an important reference for AI-SP design, while participants also described SP practice as episodic, variable, and unevenly connected to individualized review.

\paragraph{Embodied and interpersonal responses.}

Participants valued human SP practice because the encounter unfolded through visible and affective responses. Eye contact, nodding, hesitation, discomfort, and tone helped learners judge whether their questions and explanations were being received. F2 described a patient's nod as immediate feedback: \textit{``After I asked a question, the patient might nod, and that was already a kind of timely feedback.''}[F2] F4 similarly explained that a supportive nod could confirm the direction of questioning, while a hurried SP could disrupt the learner's pace: \textit{``If the SP kept nodding, I would feel that I was asking in the right direction. But if they kept rushing me, my rhythm would be disrupted.''}[F4]

These responses also mattered during physical examination. Participants described tenderness, breathing, posture, and facial expression as cues that made the encounter feel clinically consequential. F3 used abdominal examination as an example: \textit{``For tenderness and rebound tenderness, when you release your hand, the patient can show a typical expression or movement.''}[F3] Such cues informed how learners acted with the patient as well as whether they had selected the expected examination.

Participants also connected SP interaction with patients' trust in the learner. F5 recalled that when professional knowledge was explained in plain language, the SP appeared more willing to trust the learner: \textit{``I could feel that they trusted me, and that trust came from the fact that I explained professional knowledge clearly in plain language.''}[F5] At the same time, participants noted that human SPs varied in realism. Some were overly scripted, emotionally flat, or too medically fluent. Human SP practice provided embodied and interpersonal cues, but the value of those cues depended on the actor, scenario, and training arrangement.

\paragraph{Access, performance, and feedback.}

Participants often encountered human SPs through formal teaching arrangements, such as OSCEs, tutorials, or organized training sessions. F4 explained that SP access at their school was mostly tied to examinations: \textit{``At our school, we can only really encounter SPs during exams, so in daily practice we mainly rely on models.''}[F4] Participants therefore treated human SPs as valuable but not always available for repeated, self-directed rehearsal.

Feedback was similarly uneven. Some sessions included comments from SPs, senior students, or teachers, while others ended with little individualized review. F7 described an examination setting in which feedback was mostly indirect: \textit{``During the exam, I felt that I could not really receive feedback. The teacher might smile at you or sigh, but apart from that I could not feel much.''}[F7] Across these accounts, human SP practice offered bodily response and social pressure, while access and formal review depended on how the activity was organized. These findings established a design requirement for repeatable, self-directed consultation practice that retained attention to the interpersonal structure of the encounter.

\subsubsection{Extending Practice beyond Patient Dialogue}

Participants described clinical consultation as a sequence of interdependent activities that extends beyond patient conversation. Their accounts connected history taking with prior records, physical examination, auxiliary testing, diagnosis, management, documentation, and patient explanation. F1 noted that clinical reasoning could begin before direct questioning: \textit{``Before taking the history, you may first need to look at the chest image and the different test indicators.''}[F1] F11 likewise described SP practice that combined active history taking with specialist examination and outside records or imaging. These accounts positioned patient dialogue as one source of evidence within a broader clinical task.

The early prototype made some of this broader scope visible. F1 found it more comprehensive than expected because it included \textit{``examinations, laboratory tests, physical examination, and postoperative management''}[F1]. F4 similarly said that the examination and test modules resembled the clinical information systems used during patient reception. Participants therefore valued the possibility of rehearsing activities that were difficult to represent through open-ended chat alone, including selecting investigations, reviewing structured findings, and carrying evidence forward into diagnosis and management.

However, participants did not equate a larger set of modules with a broader consultation process. They identified activities that still required more detailed representation. F11 distinguished the prototype's structured clinical coverage from its limited support for complex communication, such as informed-consent conversations. Workshop participants extended this point by embedding communication within concrete clinical milestones. WS2 proposed preoperative conversations and postoperative rounds in which patients would ask questions, while WS3 suggested that learners should explain the reasons for a diagnosis to the patient. Communication was therefore treated as part of clinical progression, not as a separate conversational layer.

The workshops translated this requirement into a consultation workflow that moved from patient inquiry and clinical investigation to decision making, documentation, and patient explanation. They also wanted clinical actions to resemble familiar practice. WS1 proposed searchable test ordering because clinicians typically \textit{``enter a few keywords and then select from a large set of examinations''}[WS1]. These proposals informed the organization of MedEasy across the represented workflow.

\subsubsection{Connecting Consultation Activities across the Encounter}

Participants emphasized that adding clinical modules was insufficient unless information and actions remained connected as the case developed. Their accounts focused on sequence, persistent case state, learner agency, and consistency across dialogue and structured evidence. F11 argued that a simulated patient needed a coherent history timeline. In the workshops, WS2 made this requirement explicit: \textit{``The system needs a timeline. It cannot let you take the history, perform examinations, and order tests in an arbitrary mixture.''}[WS2] Participants did not require every case to follow an identical order, but they expected the available pathway to reflect the clinical situation, such as triage and ABCDE assessment in an emergency scenario or preoperative evaluation in a surgical case.

Visible structure also served as scaffolding. F2 explained that an entirely open interaction could be difficult because \textit{``if everything has to be explored completely on your own, it becomes harder''}[F2]. Structured options reminded learners of routine but easily omitted elements. Participants also noted that menus could replace clinical formulation. WS2 argued that urgent actions should sometimes be entered by the learner because proposing the action was itself \textit{``part of the thinking process''}[WS2]. The requirement combined workflow orientation with opportunities for learners to formulate their own next steps.

The early prototype further showed why persistent state and cross-module alignment mattered. F3 encountered a contradiction in which the patient denied hypertension during dialogue but the structured history later reported hypertension and previous surgery. The participant described such inconsistency as a central difficulty, not a superficial wording problem. Missing actions produced a related concern. When electrocardiography and other expected preoperative tests could not be found, F3 interpreted the omission through the requirements of the clinical pathway: \textit{``These are things I may need, but they were not available in the system.''}[F3] In this setting, the completeness of the available action set and the consistency of the case shaped whether learners considered the simulated consultation clinically usable.

Participants also distinguished intended clinical uncertainty from uncertainty introduced by the system. F11 noted that difficulty might arise from insufficient learner knowledge, an incomplete performance by a human SP, or an inconsistent case. The early prototype sometimes made this distinction difficult because an unanswered question could mean that the learner had phrased it incorrectly, that the speech or language system had failed to recognize it, or that the information was absent from the case. These observations motivated a clearer separation among dialogue, clinical actions, and case information, together with a record of how each changed over the course of the encounter.

The analysis produced two connected requirements: preserve relationships among patient responses, examinations, tests, decisions, and documentation, and make the available action space visible as the consultation develops.

\subsubsection{Supporting Review after Practice}

Participants did not consider final diagnostic correctness sufficient for post-practice review. They wanted to understand which information they had gathered, what they had omitted, and why a particular action or conclusion mattered. Existing SP experiences provided uneven access to this form of feedback. F10 valued sessions in which senior students discussed whether the history was complete, whether the diagnosis was appropriate, and which evidence supported or opposed it. F7 similarly described case debriefs that explained \textit{``what disease this was and why the conclusion was reached''}[F7]. These accounts emphasized process and rationale beyond a binary indication of success.

Participants also considered the timing of feedback. Several preferred a debrief after the consultation because interruption could disturb the clinical flow. F2 stated: \textit{``Do not interrupt the treatment process; tell me after I have finished.''}[F2] F11, however, recalled that immediate correction could be useful when an error needed to be repeated and repaired, while also increasing pressure and affecting subsequent performance. Feedback timing was therefore treated as contingent on its purpose: routine review could occur after the encounter, whereas severe or safety-critical errors might justify an immediate warning.

The workshops translated these preferences into proposals for reviewing the recorded encounter. WS2 argued that learners should be able to \textit{``look back at what you just did''}[WS2], linking feedback to an action history beyond an isolated score. Participants also proposed feedback through patient reactions, changes in clinical state, and textual evaluation. WS3 went further by suggesting that an incorrect diagnosis should not immediately reveal the correct answer; the system could provide additional history features and invite the learner to reason again. This proposal treated feedback as an opportunity to re-engage with the case.

These accounts motivated an action record and post-session feedback connected to case-specific criteria. Participants also emphasized that review should explain the reasoning behind omissions, selected actions, and diagnostic conclusions.

The formative findings informed four connected aspects of MedEasy: repeatable consultation practice, a workflow extending beyond patient dialogue, continuity among case information and learner actions, and post-session review of the recorded session. Table~\ref{tab:design-translation} summarizes this translation, and Section~\ref{sec:medeasysystem} describes the resulting implementation.

\begin{table}[!htb]
\centering
\caption{Translation of formative findings into MedEasy design requirements and implemented mechanisms.}
\label{tab:design-translation}
\small
\renewcommand{\arraystretch}{1.15}
\begin{tabularx}{\textwidth}{>{\raggedright\arraybackslash}p{0.23\textwidth} >{\raggedright\arraybackslash}p{0.27\textwidth} X}
\toprule
Formative finding & Design requirement & MedEasy mechanism \\
\midrule
Human SPs and real patients as reference practices & Support repeatable, self-directed consultation practice while retaining an encounter structure & Persistent case workspace, voice/text patient interaction, and reusable structured cases \\
Extending practice beyond patient dialogue & Represent clinical work across the consultation & Physical examination, auxiliary testing, diagnosis, patient education, treatment, and medical-record interfaces \\
Connecting consultation activities across the encounter & Preserve continuity among patient responses, findings, actions, and decisions & Structured case records, action-contingent findings, accumulated consultation records, and revisitable modules \\
Supporting review after practice & Relate feedback to what the learner completed and omitted & Retained consultation trajectory and post-session evaluation against expert-authored case standards \\
\bottomrule
\end{tabularx}
\end{table}

\section{MedEasy System}
\label{sec:medeasysystem}
This section describes how the design requirements in Table~\ref{tab:design-translation} were implemented in MedEasy. The system translates these requirements into a staged consultation workflow, factorized patient simulation, action-contingent clinical findings, and post-session feedback based on the retained consultation trajectory.

\begin{figure}[!htb]
\centering
\includegraphics[width=\linewidth]{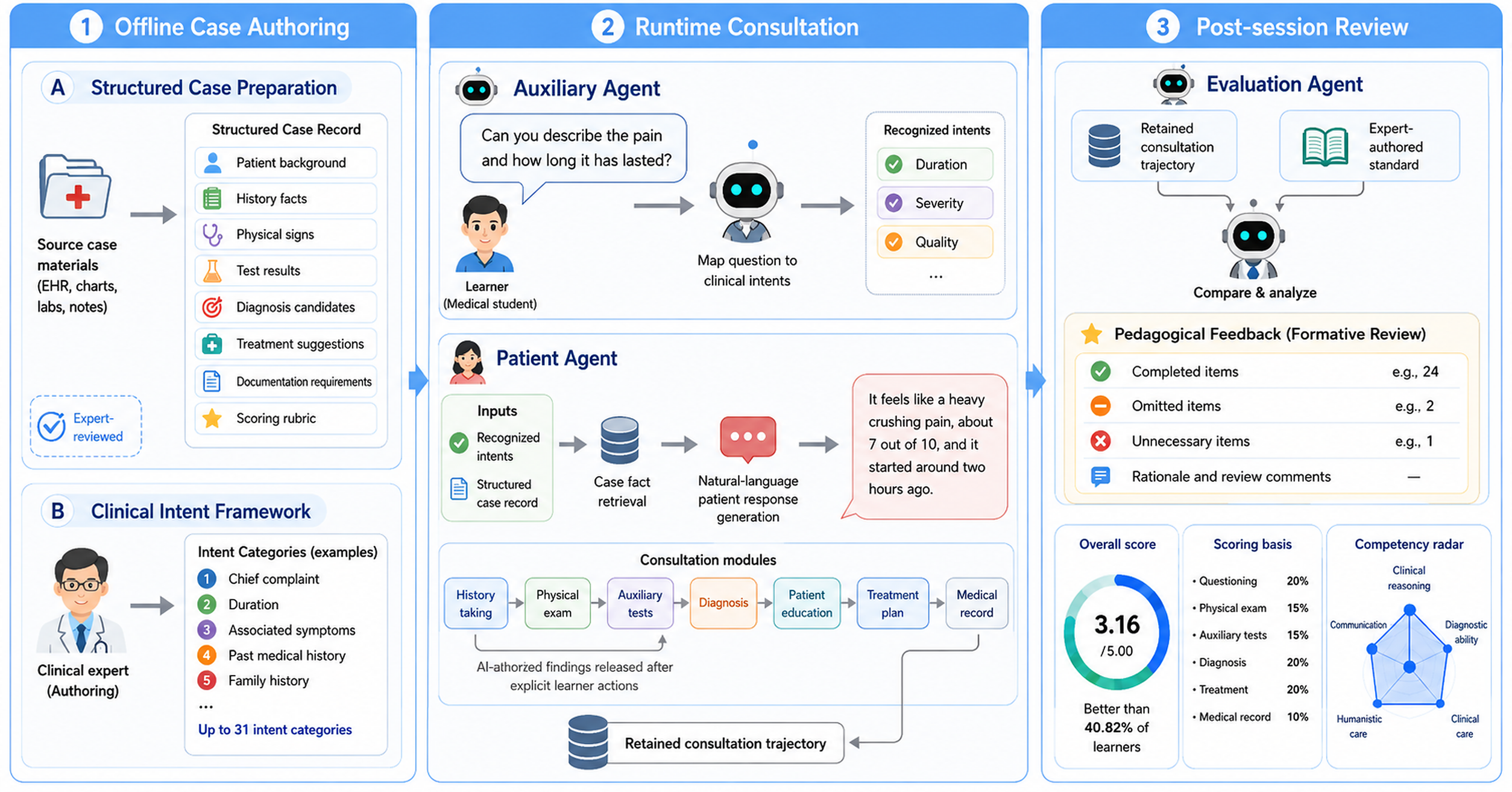}
\caption{MedEasy system architecture and data flow. Offline case authoring produces structured case records and a clinical intent framework. During runtime consultation, the Auxiliary Agent maps learner utterances to clinical intents, the Patient Agent generates case-grounded responses, and clinical-action interfaces release authored findings after explicit learner actions. The retained consultation trajectory is then reviewed by the Evaluation Agent against expert-authored standards for formative post-session feedback.}
\label{fig:system-overview}
\end{figure}

MedEasy is a web-based, multi-agent virtual standardized patient system for structured consultation practice. The system treats a virtual consultation as a staged process that includes patient dialogue, clinical action, decision submission, documentation, and post-session review. Its architecture separates intent recognition, case-grounded patient response, action-contingent clinical findings, and trajectory-level evaluation. This separation follows the implemented EasyMED framework, in which the Auxiliary Agent identifies the learner's clinical inquiry, the Patient Agent generates the patient response from structured case information, and the Evaluation Agent reviews the completed session against expert criteria \citep{zhang2026easymed}. Figure~\ref{fig:system-overview} summarizes the system architecture and data flow. The complete prompts and additional technical reference tables are provided in Appendices B and C.

\subsection{Consultation Workflow}
\label{sec:workflow}

\begin{figure}[!htb]
\centering
\includegraphics[width=\linewidth]{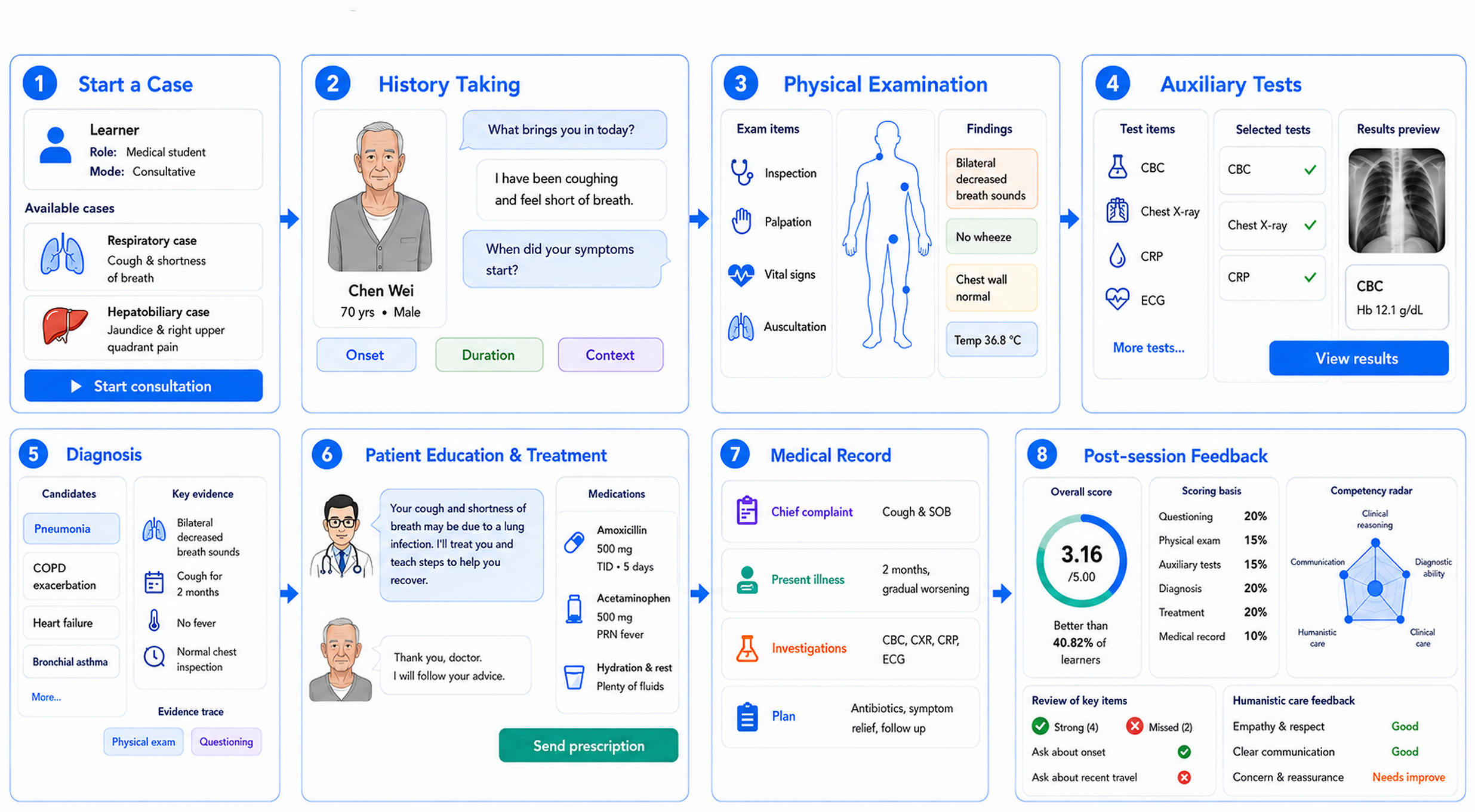}
\caption{Overview of the MedEasy consultation workflow. Learners move from patient questioning to physical examination, auxiliary testing, diagnosis, patient education, treatment planning, medical-record writing, and post-session feedback. The figure shows the learner-facing modules through which MedEasy organizes consultation practice as a staged workflow.}
\label{fig:consultation-workflow}
\end{figure}

A MedEasy session begins in the main dialogue interface. The interface includes an information and control panel, a three-dimensional virtual patient avatar, and an interactive chat module. Learners conduct the medical history interview through voice or text. The system retains the preceding doctor--patient dialogue so that later turns can be interpreted in context. A persistent consultation record allows learners to review the developing case as the encounter proceeds. Figure~\ref{fig:consultation-workflow} shows the learner-facing workflow across these stages.

\begin{figure}[!htb]
\centering
\includegraphics[width=\linewidth]{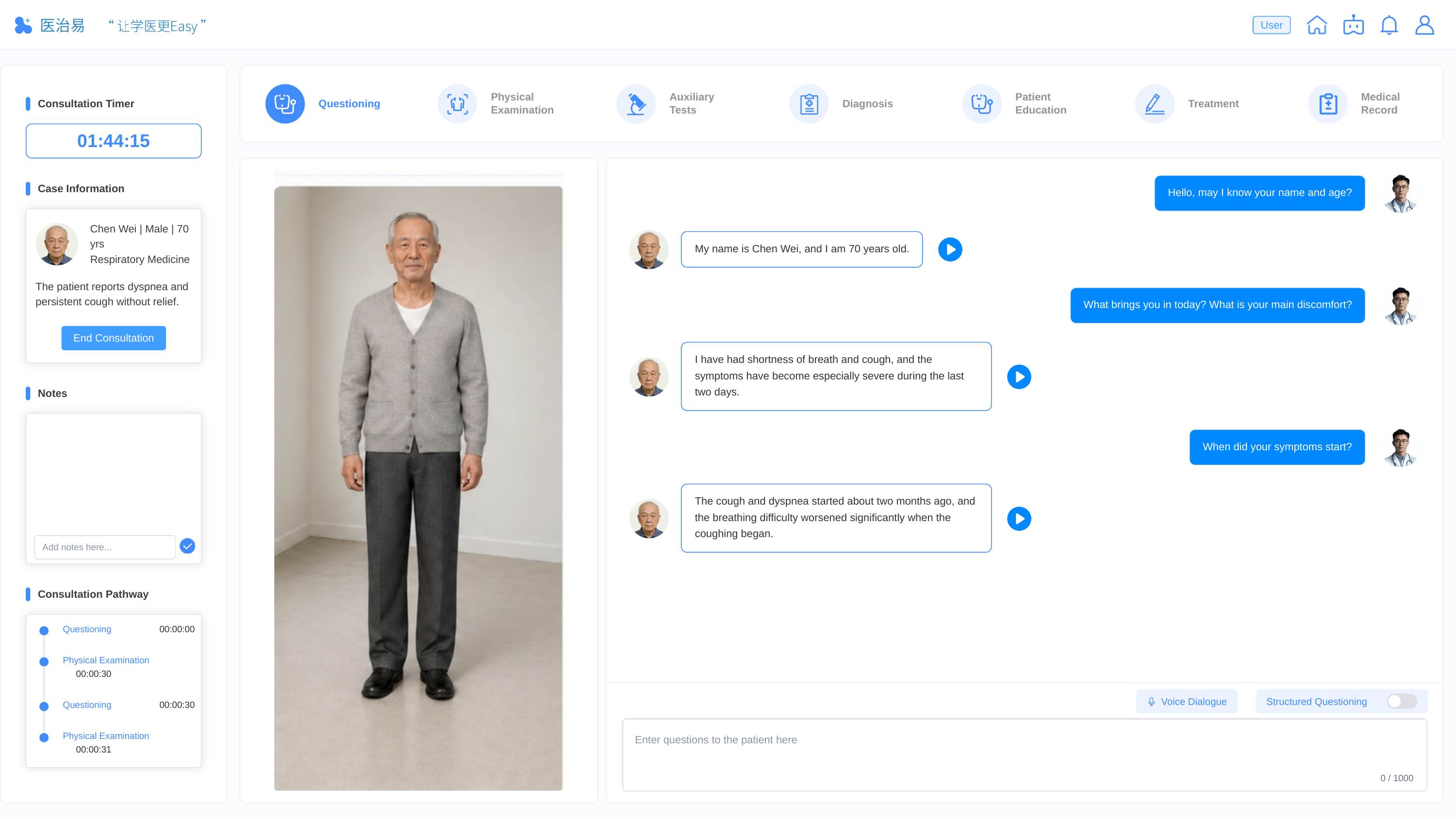}
\caption{Questioning interface in MedEasy. The interface combines patient dialogue, case information, consultation notes, voice/text input, and the consultation pathway. Learners use this page to conduct history taking while the system records the dialogue context for later interpretation and review.}
\label{fig:questioning-interface}
\end{figure}

During history taking, each learner utterance is first processed by the Auxiliary Agent. The agent maps the utterance to one or more predefined clinical intents. The Patient Agent then receives the corresponding case information and generates a patient response. This workflow separates the selection of clinical information from the natural-language expression of that information. Figure~\ref{fig:questioning-interface} shows the questioning interface used during this stage.

\begin{figure}[!htb]
\centering
\includegraphics[width=\linewidth]{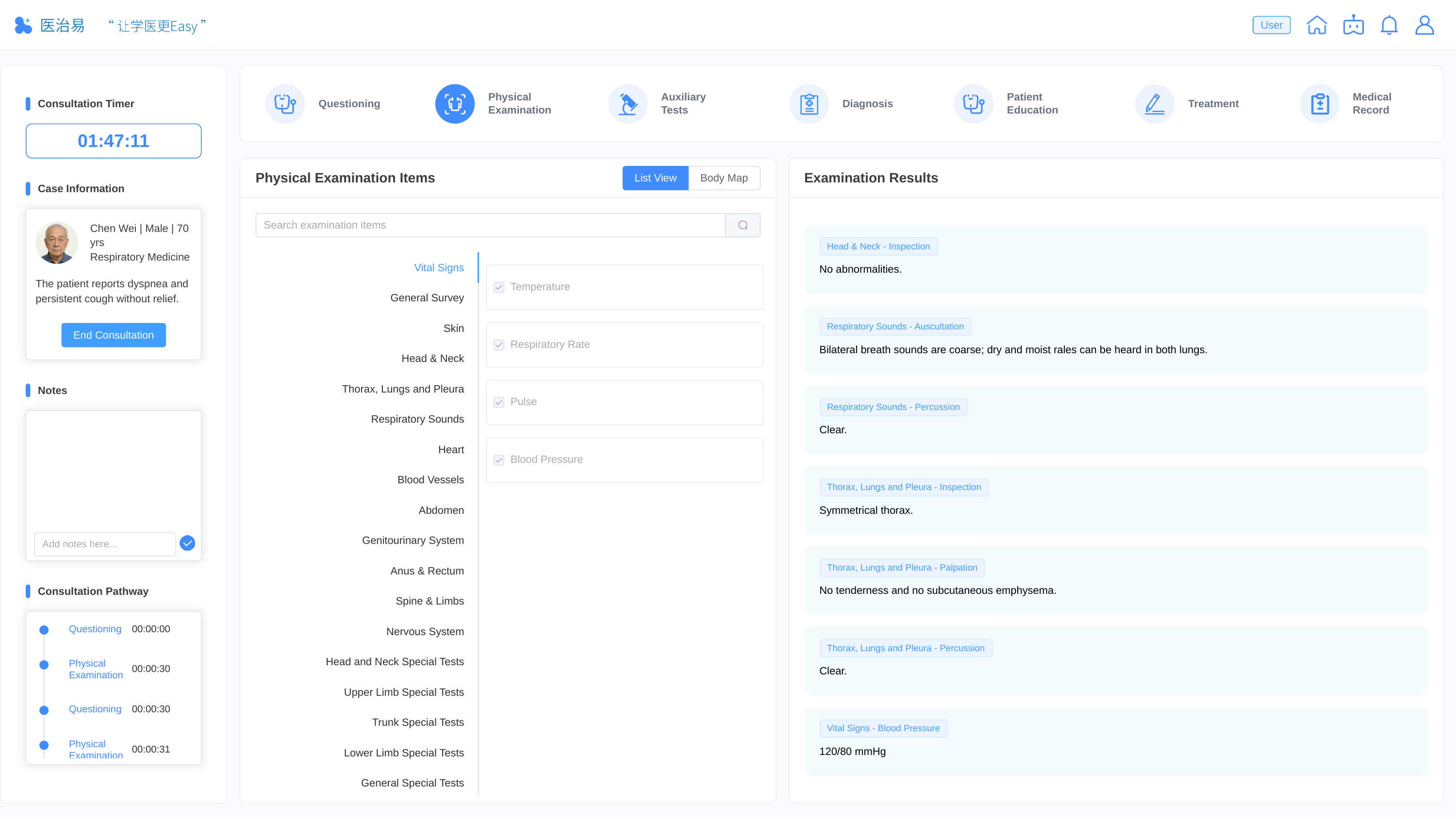}
\caption{Physical-examination interface in MedEasy. Learners select body regions and examination actions through the anatomical interface. The system then displays authored examination findings in the result panel, keeping physical-examination information separate from generated patient dialogue.}
\label{fig:physical-examination-interface}
\end{figure}

After the dialogue stage, learners continue through clinical-action interfaces. Physical examination is performed through an anatomical interface: learners select a body region and an examination technique, and the corresponding authored finding is displayed in the result panel. Auxiliary examinations are ordered through a searchable list of laboratory and imaging tests. Learners add selected tests to a request queue and review the returned results. In the diagnosis interface, learners inspect the accumulated case information and submit a diagnosis. Figure~\ref{fig:physical-examination-interface} illustrates the physical-examination interface.

The consultation then continues through patient education, treatment planning, and medical-record writing. Learners submit a management plan in the treatment interface. In the medical-record interface, they enter structured free-text documentation for the case. The interface retains the consultation pathway and allows learners to return to earlier modules before ending the consultation.

After the learner ends the case, the post-session evaluation interface presents the overall performance result, consultation duration, patient information, and dialogue timeline. It aligns recorded doctor--patient exchanges with relevant clinical objectives and provides itemized feedback on completed and omitted steps.

\subsection{Factorized Patient Simulation}
\label{sec:patient-simulation}

\paragraph{Auxiliary Agent.}
The Auxiliary Agent models the learner's clinical inquiry before a patient response is generated. For each learner question, it maps surface-level phrasing to a predefined clinical intent. The intent label is then used to select the case information supplied to the Patient Agent.

The intent framework contains 31 categories covering the scope of a medical history interview, including the chief complaint, present illness, system review, medical history, personal and family history, and patient-centered communication. Each utterance may be assigned up to three intents. The agent uses the conversation history to interpret vague or context-dependent inputs, returns intent labels without explanation, and defaults to the small-talk category when an intent cannot be confidently determined. Appendix C lists the complete intent framework, while Appendix B reproduces the classification prompt.

\paragraph{Patient Agent.}
The Patient Agent formulates patient responses under case-information and disclosure constraints. Given an inferred clinical intent, it retrieves the corresponding fact from the structured case record and generates a natural-language response conditioned on that fact, the patient persona, the purpose of the consultation, and the conversation history.

The Patient Agent is instructed to follow the supplied case information, avoid unsupported additions, translate medical terminology into lay language, maintain a concise and emotionally consistent patient voice, and answer only what the learner asks. It uses the conversation history to resolve references and maintain continuity. The prompt also includes role-selection rules for pediatric cases, instructions about system identity, responses to broad requests for medical history, and guidance for handling inappropriate language. Appendix B reproduces the complete prompt as used in the system.

\subsection{Structured Case Library and Action-Based Result Release}
\label{sec:case-representation}

The MedEasy case library contains 20 clinical cases selected by a panel of medical experts from the \textit{Peking Union Medical College Hospital Clinical Thinking Training Case Collection}. The cases were selected to align with the curriculum for fifth-year undergraduate medical students. The library is balanced by patient gender and spans three age groups, seven major organ systems, and 16 diagnoses; Appendix C provides an overview of its distribution.

To make the source cases compatible with the multi-agent workflow, medical experts defined a structured case template covering patient background, clinical history, physical signs, test results, and emotional tone. The original narrative cases were first converted into the template using GPT-4o with task-specific prompts. Medical experts then reviewed and corrected every generated field before the cases were added to the system.

\begin{figure}[!htb]
\centering
\includegraphics[width=\linewidth]{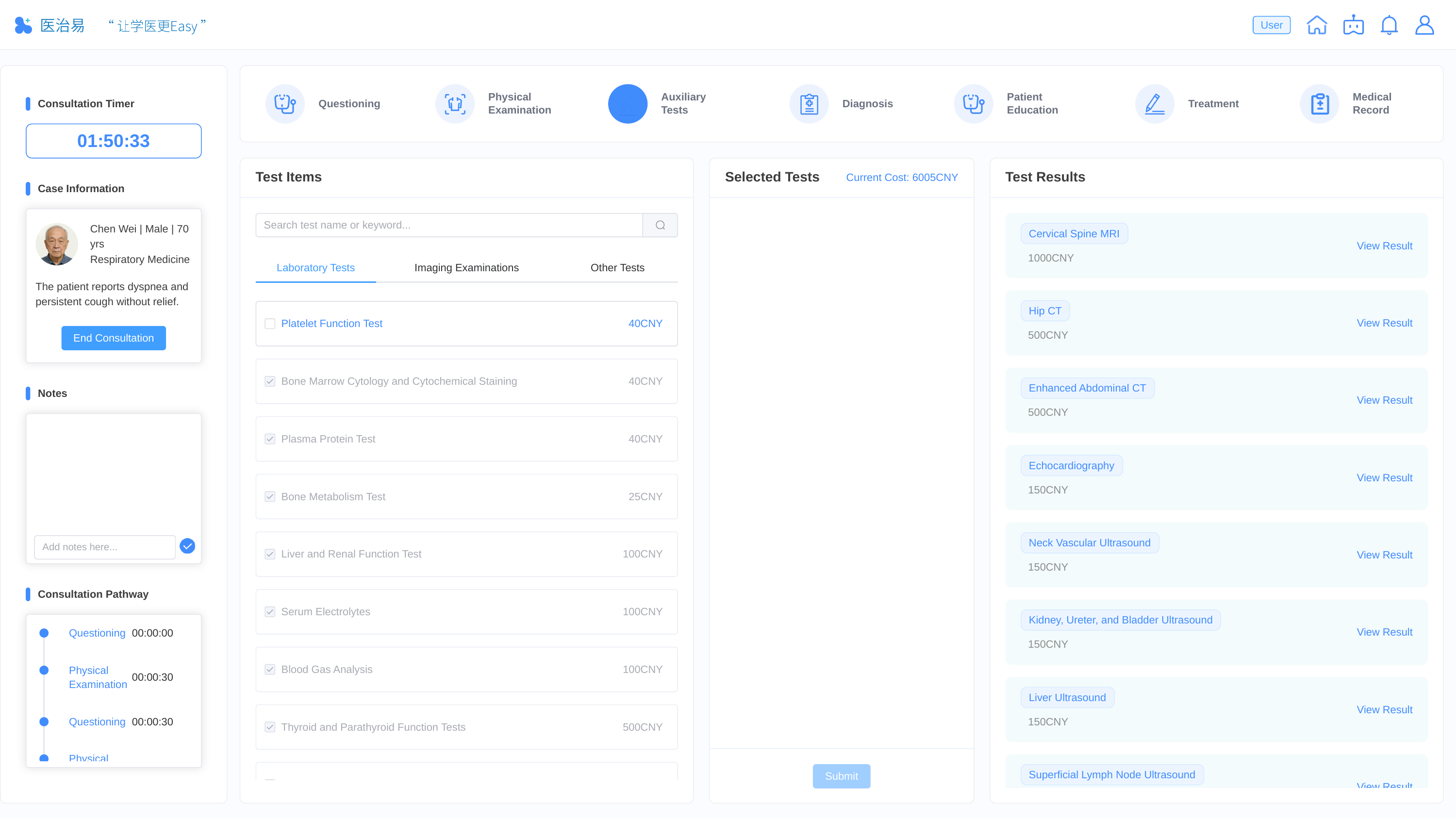}
\caption{Auxiliary-test ordering interface in MedEasy. Learners search for laboratory and imaging tests, add selected tests to a request queue, and review returned results. Test findings are retrieved from the structured case record after explicit learner actions.}
\label{fig:auxiliary-tests-interface}
\end{figure}

\begin{figure}[!htb]
\centering
\includegraphics[width=\linewidth]{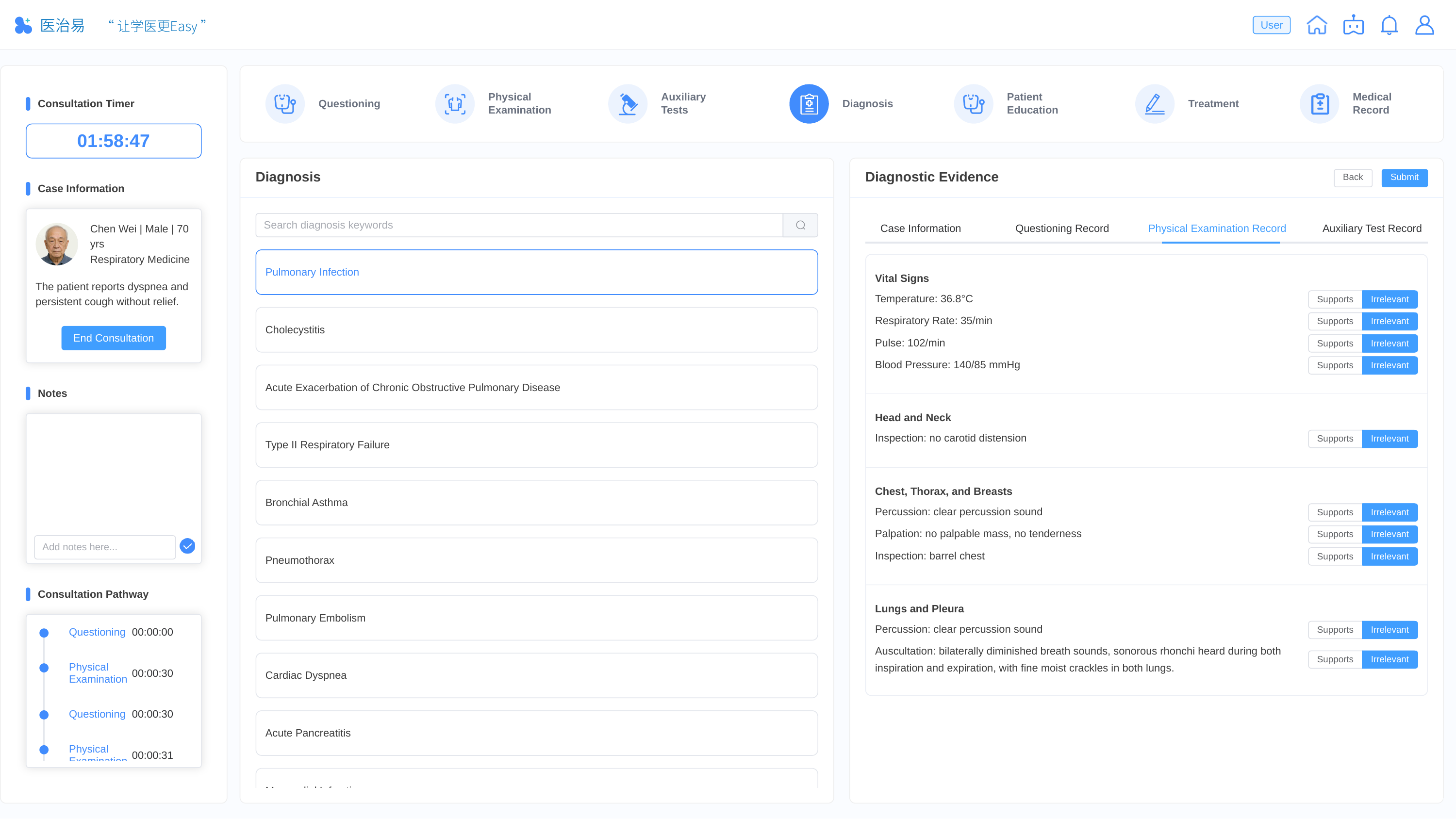}
\caption{Diagnosis interface in MedEasy. The interface brings together accumulated evidence from prior questioning, physical examination, and auxiliary testing. The evidence trace links case information and clinical actions to diagnostic reasoning, helping learners inspect how the represented case supports or weakens candidate diagnoses.}
\label{fig:diagnosis-interface}
\end{figure}

The structured case record is the shared source for patient dialogue, clinical-action interfaces, and post-session evaluation. During dialogue, the Patient Agent receives only the case information associated with the recognized intent. Physical-examination and auxiliary-examination findings are returned from the structured record after the learner selects the corresponding action. The language model is not used to invent examination or test findings. Figure~\ref{fig:auxiliary-tests-interface} shows the auxiliary-test ordering interface, and Figure~\ref{fig:diagnosis-interface} shows how accumulated evidence is represented during diagnosis.

As the consultation develops, the system retains the dialogue, recognized intents, elicited case facts, selected clinical actions, submitted decisions, and medical-record entries. This retained trajectory is passed to the Evaluation Agent after the session ends.

\subsection{Post-Session Evaluation and Feedback}
\label{sec:case-feedback}

The Evaluation Agent is a post-session evaluation component. It does not intervene during the consultation. After the learner ends the case, it receives the student performance record and the expert-authored reference answer. The performance record includes the dialogue history, recognized intents, elicited facts, clinical actions, diagnostic submission, treatment plan, and medical-record entry.

\begin{figure}[!htb]
\centering
\includegraphics[width=\linewidth]{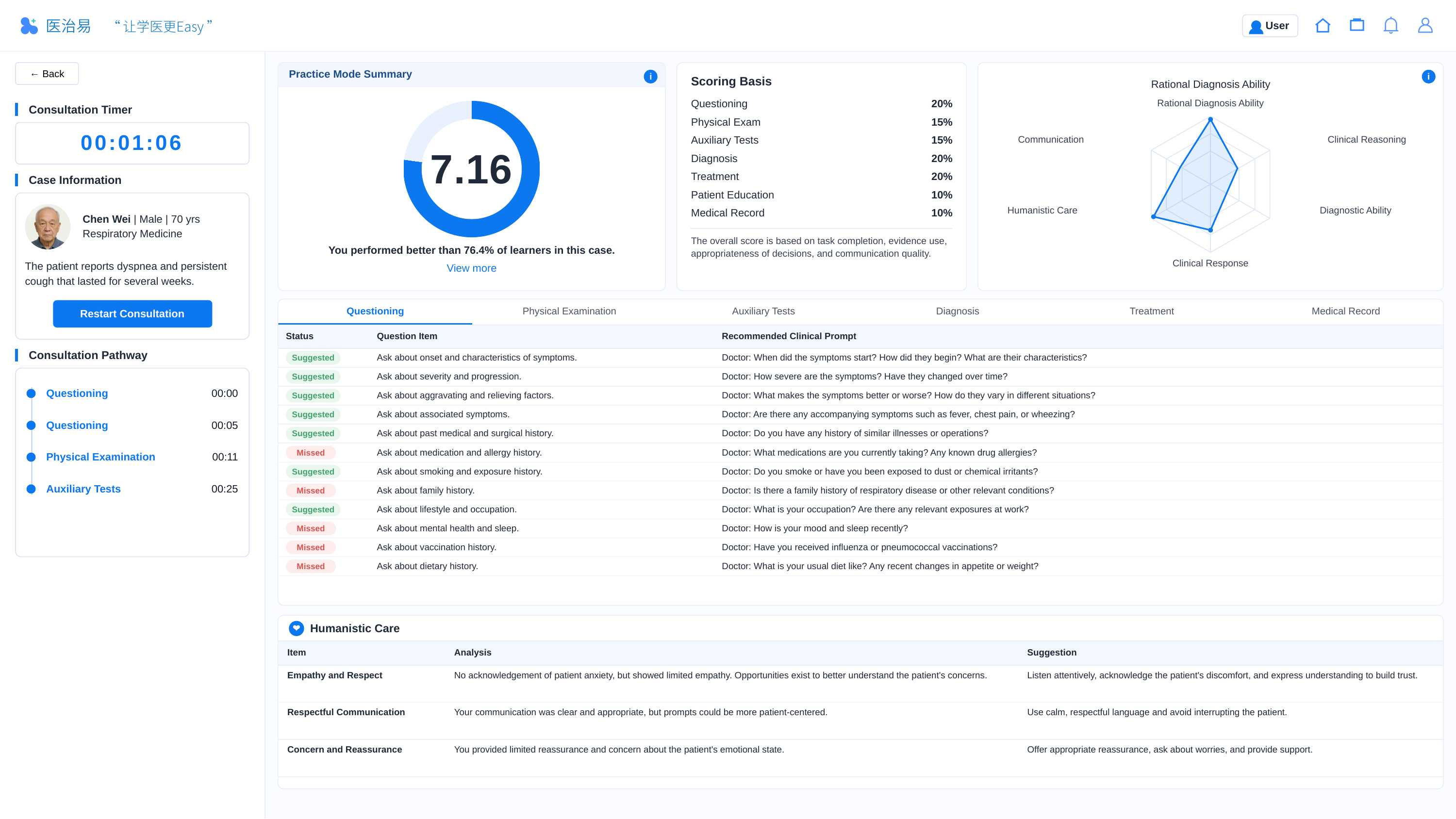}
\caption{Post-session feedback interface in MedEasy. After the consultation, the system presents module-level performance, scoring basis, competency summary, itemized review, and humanistic-care feedback. The review is generated from the retained consultation trajectory and expert-authored case standards and is used for formative learning rather than formal grading.}
\label{fig:post-session-feedback}
\end{figure}

The evaluation follows six modules: history taking, physical examination, auxiliary examination, diagnostic reasoning, treatment planning, and overall performance. For each module, the agent compares the recorded activity with the expert answer, identifies completed and omitted items, and explains the diagnostic relevance of omissions when applicable. Its core instructions prohibit adding requirements outside the expert standard, penalizing items absent from the standard, or evaluating content beyond the supplied answer. The generated review applies the scope defined in the authored teaching case and is used for formative post-session review. Figure~\ref{fig:post-session-feedback} shows the post-session review interface. Appendix B reproduces the complete Evaluation Agent prompt.

\subsection{Implementation}
\label{sec:implementation}

MedEasy uses a web-based client--server architecture. The frontend presents the consultation modules and the post-session evaluation interface. The backend orchestrates the Auxiliary, Patient, and Evaluation Agents, retrieves structured case information, records the consultation trajectory, and prepares the session summary for evaluation.

The Patient Agent and Evaluation Agent use Gemini 2.5 Pro, while the Auxiliary Agent uses Gemini 2.5 Flash. The implementation retains the available dialogue or session context within a configured 256k-token context window. All agents use a temperature of 0.7. Sessions have no fixed turn limit and end when the learner completes the case or chooses to stop.
\FloatBarrier
\section{Evaluative User Study}
\label{sec:user-study}

The evaluative user study took place between March and May 2026 and involved a cohort separate from the formative study. It examined how learners interpreted and used the implemented MedEasy system. Sessions and interviews were conducted in Mandarin Chinese, audio-recorded, transcribed, and de-identified. Coding was performed on the original Chinese-language transcripts. Quotations selected for publication were translated into English and checked against the source transcripts. We identify evaluative user-study participants as P1--P12.

\subsection{Participants}

We recruited 12 clinical-year medical students. None had participated in the formative interviews or workshops. All had prior experience with human SP encounters or closely related clinical communication and OSCE training. They also had prior experience with AI-mediated patient tools, virtual patients, symptom checkers, or comparable conversational clinical learning tools. Recruitment took place through institutional mailing lists, student communities, and snowball sampling. Participation was voluntary, unrelated to academic assessment, and compensated. Participant characteristics are reported in Table~\ref{tab:participant-information}(b).

\subsection{Study Cases}

We selected two cases from the MedEasy clinical case library. One was a respiratory-medicine case involving chest pain and fever. The other was a hepatobiliary--pancreatic surgery case involving abdominal symptoms and observable signs. Each case used the structured representation, clinical-action interfaces, and expert standard-answer format described in Section~\ref{sec:case-representation}. The two cases were chosen to provide different clinical presentations and different relationships among patient-reported information, examination findings, auxiliary examinations, diagnosis, and treatment planning.

\subsection{Procedure}

Each participant completed an individual study session lasting approximately 90--120 minutes. After providing informed consent, participants received a brief introduction to MedEasy and completed both cases using the system. Case order was counterbalanced across participants.

Participants completed the full MedEasy workflow for each case, from patient questioning to post-session feedback. They could return to earlier modules, revisit previously obtained information, and revise or verify an action. Participants were asked to approach each case as they would in clinical training and to use the available functions according to their own judgment.

Following system use, we conducted a semi-structured interview. Participants described how they approached the consultation, which information they considered clinically meaningful, and how the interface shaped their actions. We asked about moments when the system seemed consistent, incomplete, or questionable, and how participants proceeded when uncertainty arose.

The interview also invited participants to reflect on MedEasy in relation to prior human SP encounters, virtual-patient systems, AI-mediated patient tools, and clinical training. These prior experiences served as points of reference and were not standardized study conditions. The session concluded with questions about which parts of consultation training continued to require human SPs, clinical educators, physical models, or real patients. Interviews were audio-recorded, transcribed, and de-identified.

\subsection{Evaluative User-Study Data Analysis}

We analyzed the evaluative user-study interviews as a separate dataset, following the same team-based deductive–inductive hybrid thematic analysis procedure described in Section \ref{sec:formative_analysis}. The formative study data analysis focused on deriving design requirements, while this evaluative study data analysis examined how 12 different medical students used and interpreted the implemented MedEasy system.

The initial coding categories in the code manual were adapted to the evaluative focus and covered consultation workflow, case representation, clinical action, system breakdowns, feedback interpretation, and MedEasy's role in clinical training. Relevant excerpts were converted into statement cards using the same format as in Section \ref{sec:formative_analysis}. This process produced 239 statement cards.

The researchers then compared and grouped the cards through affinity diagramming \citep{lucero2015affinity}, resulting in four evaluative themes. As in the formative analysis, we used consensus-based coding rather than calculating inter-rater reliability, and disagreements were resolved through discussion.

After the four evaluative themes had been developed, we related them back to the formative design requirements to clarify how the implemented system was interpreted in use. The four themes concerning the implemented system are presented in the following section, and the 12 participants are labeled as P1–P12.

\section{Evaluative User-Study Findings}

The evaluative user-study analysis produced four themes. Participants described MedEasy as a multi-stage consultation environment in which patient questioning was connected with examination, testing, diagnosis, management, documentation, and review. They used prior simulation and AI-mediated learning experiences to interpret this form of practice. Their accounts also showed how explicit system structure made the relationships among responses, findings, actions, and criteria available for inspection, while leaving the source of some inconsistencies unclear.

\subsection{Theme 1: Organizing Information and Decisions across the Consultation}

Participants described prior AI-mediated patient activities as commonly organized around conversational clue gathering and diagnosis identification. MedEasy presented a broader sequence in which information obtained in one stage remained relevant to later examinations, decisions, and documentation. Visible modules gave participants a route through the case, while also defining the actions through which the system represented consultation practice.

\noindent\textbf{Diagnosis-Oriented AI Practice as a Reference Point.} Participants recalled virtual-patient systems, symptom-checking applications, and conversational AI tools that focused on symptom inquiry or diagnosis identification. P6 said that some activities ended when \,\textit{``once you guessed the disease, it was basically over.''}[P6] P7 described a similar pattern: \textit{``you just keep asking the patient and let the patient give you the answer.''}[P7] In these accounts, the learner progressed by collecting enough conversational clues to identify a disease.

Participants did not dismiss this format. P3 said that diagnosis-oriented systems \textit{``also have some help for medical students' learning''}[P3], and P8 associated them with \textit{``quick differential diagnosis''}[P8]. P4, however, recalled that one system felt \textit{``more like a puzzle game''}[P4]. These prior experiences gave participants a reference for interpreting MedEasy: dialogue remained important, but the represented task continued after a likely diagnosis had emerged.

\noindent\textbf{Moving through a Multi-Stage Case.} Participants described history taking as a progressive inquiry. P1 explained, \textit{``You first ask what symptoms the patient has, and then slowly go deeper.''}[P1] P10 associated the sequence with diagnostic medicine training, describing it as \textit{``more like the course of clinical diagnostics, following those steps.''}[P10] Their accounts treated the dialogue as the beginning of a developing case record.

Examination and test modules introduced new information through explicit actions. P5 recalled, \textit{``After I finished the examination, it would directly give me the result.''}[P5] Participants then described using these findings when considering later parts of the case. P8 characterized the sequence as a \textit{``hypothesis and verification process of differential diagnosis''}[P8]. P3 contrasted this with prior activities in which \textit{``there is no treatment content and no medical record writing''}[P3]. Across these accounts, the represented consultation required information gathered earlier to remain available when participants described diagnosis, management, and documentation.

\noindent\textbf{Visible Structure as Guidance and Constraint.} Visible options shaped how participants described navigating the case. P3 noted that when examination options were displayed, \textit{``it is easier to think of them''}[P3]. P2 considered the explicit sequence suitable for learners who were still becoming familiar with consultation routines. The interface therefore acted as a reminder of possible next steps.

The same structure defined what the system could recognize and record. Participants could select only the examinations, tests, and treatments represented in the interface. P1 valued the clear sequence, while P10 described the activity as more formal than some previous AI-mediated exercises. These accounts also qualified the apparent completeness of the workflow: the displayed path represented the consultation designed into the system, not every clinically defensible action. In response to RQ1, participants interpreted MedEasy as practice in organizing information and decisions across several consultation stages, using diagnosis-oriented tools as a narrower point of comparison.

\subsection{Theme 2: Locating the Source of Inconsistency and Breakdown}

Participants could inspect generated patient responses, authored findings, available actions, and later feedback, but they had limited visibility into how the system had interpreted a question or recorded an action. When information did not align, they described comparing other parts of the case and trying to determine whether the difficulty came from their own reasoning, the authored case, or the interface.

\noindent\textbf{Clinical Framing and Initial Confidence.} Participants often formed an initial impression from the clinical framing of the system. P7 said that MedEasy \textit{``felt more like a hospital system''}[P7]. P6 similarly noted, \textit{``It tells you which department this case belongs to, so it feels more like a real clinical situation.''}[P6] Department labels, examination options, and test modules made the activity appear more formal than a general chatbot.

Participants then compared the case with their own clinical knowledge. P5 explained, \textit{``If the examination result matches what I expected, I will feel that this case is relatively reliable.''}[P5] P2 described confidence when symptoms, findings, and diagnosis followed patterns learned in class. P8 nevertheless distinguished appearance from correctness: \textit{``At first I felt it was quite standard, but I still had to see whether the following examination and tests could match it.''}[P8] Clinical presentation shaped the starting point of judgment, but later information could qualify that impression.

\noindent\textbf{Cross-Component Checking.} Participants described checking whether dialogue, examination findings, test results, and final conclusions corresponded. P8 said, \textit{``I would compare what the patient said with the examination and test results.''}[P8] P9 noted that feedback would be difficult to accept if it referred to evidence that had not appeared earlier. P11 described a coherent case in which \textit{``the previous information and the final result could correspond to each other.''}[P11]

P8 provided the clearest account of responding to a questionable answer. They recalled reconsidering their own assumption and consulting other evidence: \textit{``At that moment I wondered whether my own judgment was biased, so I went back to look at other evidence.''}[P8] They also described rephrasing a question and then checking later findings: \textit{``I changed the way I asked, and then I looked at whether the later examination could support it.''}[P8] This account shows one way participants described working around uncertainty, but it was not a uniform strategy across the cohort.

\noindent\textbf{Missing Actions and Represented Case Coverage.} Participants also evaluated the actions absent from the interface. P5 recalled, \textit{``I wanted to choose that examination, but there was no such option.''}[P5] P6 described a similar problem when a common test could not be found. P11 connected the absence of a treatment option to uncertainty about their own reasoning: \textit{``If there is no medicine I think should be used, I do not know whether I am wrong or the system did not include it.''}[P11]

Participants did not expect every possible action to be represented. P8 argued that the system should cover actions relevant to the case, while P12 considered omissions acceptable when they fell outside the intended focus. Their concern centered on missing options that appeared necessary for a plausible pathway. The action space therefore became part of how they judged the scope of the represented case.

\noindent\textbf{Technical and Procedural Uncertainty.} Delays, login problems, unclear controls, and speech-recognition errors interrupted the activity. P7 said, \textit{``If it is stuck for a long time, I will start to doubt whether the system can continue the case.''}[P7] P12 described voice-recognition problems that made it difficult to know whether an utterance had been captured.

Other uncertainties concerned the workflow itself. P4 recalled, \textit{``Sometimes I was not sure what I should do next.''}[P4] P12 described being unable to tell whether the system had failed to recognize an action or whether the question had been phrased incorrectly. These accounts answer RQ2 by showing that participants assessed relationships across system components, but often lacked enough process visibility to identify where a breakdown had occurred.

\subsection{Theme 3: Reviewing Performance through Case-Specific Criteria}

Participants read post-session feedback in two ways. It reconstructed parts of the consultation by showing recognized actions and omissions, and it applied a case-specific account of what should have been done. Participants valued the recorded review while questioning criteria that did not fit the specialty context, the information available at the time of action, or practical constraints on care.

\noindent\textbf{Reconstructing the Recorded Consultation.} P2 described the review as showing \textit{``what I had done in each part, and what I had omitted''}[P2]. P1 similarly said that the final evaluation identified \textit{``which parts were judged accurately and which parts were wrong''}[P1]. These accounts referred to a record extending beyond the final diagnosis. Participants could inspect how the system represented the preceding consultation and which parts entered the evaluation.

P9 valued feedback that separated completed, unnecessary, and missing actions: \textit{``which things you did were correct, which were unnecessary, and which you did not do''}[P9]. P7 contrasted this with prior diagnosis-oriented activities that lacked \textit{``a feedback page telling you which areas needed improvement''}[P7]. The review gave participants a basis for revisiting the process recognized by the system.

\noindent\textbf{Questioning the Applied Criterion.} Explanations became important when a comment identified an omission or rejected an action. P8 valued comments that described the \textit{``clinical significance''}[P8] of a missing item. P5, however, questioned why gastrointestinal and bowel questions were required in a respiratory case: \textit{``He was registered in the respiratory department, so I did not really understand why not asking this was considered wrong.''}[P5] P10 similarly wanted to know \textit{``why this disease does not require this examination''}[P10] when an imaging action was classified as unnecessary.

These comments did not only concern wording. Participants were examining the standard used to classify an action. A feedback statement could be traceable to the recorded session and still remain disputable as a clinical judgment.

\noindent\textbf{Reading Standards in Context.} P6 distinguished textbook completeness from actual consultation practice: \textit{``doctors do not ask this completely in actual consultations''}[P6]. P9 made a similar point: \textit{``In clinical practice, you simply cannot ask that many things.''}[P9] Participants assessed whether a completeness requirement fit the available time and the apparent specialty scope.

Cost and patient burden also entered these judgments. P6 noted that \textit{``the patient may not be willing to spend that much money on tests''}[P6], and P8 suggested displaying approximate costs. P8 also disputed the retrospective classification of a nonspecific infection test as irrelevant. Before the diagnosis had been established, the test could have been used to examine an alternative. This account showed that the relevance of an action depended on the learner's information state at the time it was chosen.

\noindent\textbf{Standards as Material for Further Review.} Participants still valued an explicit reference. P10 argued that medical learning required \textit{``a standard answer''}[P10], and P6 valued feedback on the completeness of history taking and examination. They did not, however, treat the displayed standard as self-validating. Disputed comments prompted questions about why the criterion applied and whether an alternative pathway had been recognized.

Feedback concerning interpersonal communication appeared less often in participants' accounts than comments on evidence, examinations, and diagnostic reasoning. P8 recalled a comment about insufficient concern for the patient, but most discussion focused on discrete actions and omissions. In response to RQ2, participants interpreted feedback through both its connection to the recorded trajectory and the contextual legitimacy of the criterion it applied.

\subsection{Theme 4: Positioning AI-SP Practice among Other Training Formats}

Participants located MedEasy within a broader set of training resources. They described the system as suitable for repeatable practice in organizing a represented consultation, while assigning embodied response, procedural contact, instructional interpretation, and responsibility for patient welfare to other formats.

\noindent\textbf{Repeatable Rehearsal before Situated Encounters.} P4 called MedEasy \textit{``a transitional stage between just learning something and going into the clinic''}[P4]. P1 associated it with \textit{``standardized and procedural learning''}[P1], and P5 described using the case to rehearse which examinations should be considered. Participants could repeat a case and revisit earlier modules without the scheduling and interpersonal pressure of a human SP session.

They also qualified this role. P1 noted that \textit{``some rare situations may not be fed back very well''}[P1]. P8 described the activity as close to clerkship rehearsal while distinguishing it from patient care. Participants most often positioned the system as preparation for learners who were still forming a consultation routine, not as a substitute for later situated practice.

\noindent\textbf{Idealized Patient Interaction.} Participants described the AI patient as clearer and more cooperative than many real patients. P4 said, \textit{``real patients will not answer whatever you ask so clearly''}[P4], while P5 called the AI patient \textit{``a very ideal patient''}[P5]. P9 also noted that the patient sometimes used professional terminology and was \textit{``not very close to reality''}[P9].

P5 contrasted this interaction with patients who use dialects, rely on family members, answer indirectly, or withhold information. Such encounters require learners to respond to ambiguity, resistance, emotion, and competing accounts. Participants therefore described MedEasy as concentrating on the organization of case information under comparatively cooperative interaction conditions.

\noindent\textbf{Embodied and Procedural Work.} Participants separated receiving an examination finding from learning how to examine a person. P3 referred to \textit{``inspection, palpation, percussion, and auscultation''}[P3]. P4 described the AI-based version as follows: \textit{``For physical examination, with AI you just open an examination and read a report.''}[P4] The system represented the finding in the case, but not the touch, pressure, body position, or patient reaction involved in producing it.

P8 made the same distinction for suturing, puncture, and intubation: \textit{``AI has no way to do that.''}[P8] Participants discussed VR, physical models, and sensors as possible additions, but continued to treat procedural contact as a separate form of training work.

\noindent\textbf{Human and Instructional Roles.} Participants assigned technique demonstration and interpretation of disputed standards to educators. P8 said, \textit{``For physical examination, the methods for different parts still need a teacher to demonstrate.''}[P8] Human SPs contributed bodily and interpersonal response; P2 said that examination still required \textit{``a real person to react''}[P2]. Physical models represented contact with instruments and tissue, while clinical encounters added uncertainty and consequences for patient welfare.

P12 described reaching an initial symptom and then not knowing how to continue: \textit{``Sometimes, after the first symptom, I do not know what to do next.''}[P12] For such learners, an AI-SP could provide a repeatable sequence before supervised or embodied practice. In response to RQ1, participants did not place the available formats on a single scale of realism. They differentiated them by the form of consultation, interaction, procedure, interpretation, or responsibility each one represented.

\section{Discussion}

The evaluative user study shows that learners interpreted MedEasy through the consultation as it unfolded. The AI patient's responses were important, but participants read them together with later parts of the session as they moved from dialogue into clinical work and review. Their understanding of the system developed through use. This pattern is consistent with accounts of trust in automation and AI as an experience-based assessment of system behavior \citep{lee2004trust,mayer1995integrative,hoff2015trust}. In MedEasy, that experience centered on whether the system could maintain one coherent clinical case across the session.

This finding also clarifies the role of the formative requirements. The formative study had identified the need for repeated consultation practice, clinical work beyond patient dialogue, continuity across the encounter, and post-session review. The evaluative study showed how these requirements appeared in use. Repeated practice became a way for participants to organize their consultation routine and revisit weak points. Clinical work beyond dialogue became visible as participants moved from patient interaction to examinations, decisions, documentation, and review. Continuity became harder to manage when unclear responses, missing actions, or unexpected feedback made participants question whether the issue came from their own reasoning or from the system. Post-session review became valuable when feedback pointed to completed and omitted actions, and more difficult when participants questioned the case standard behind that feedback.

Together, the two phases show that the main design issue is how an AI-supported training system holds a clinical case together across use. In MedEasy, generated dialogue was only one part of the consultation. Learners also selected actions, documented decisions, and reviewed feedback against a case standard. The following sections discuss how this pattern reshapes trust, case design, feedback, and the role of AI simulation in professional training.

\subsection{Trust and Legibility in Structured Consultation}

Participants often formed their first judgment of MedEasy through patient dialogue. A response that sounded clinically plausible gave them a basis for the next clinical action. This confidence remained provisional. Learners later checked whether the same case still made sense when findings and feedback appeared. A patient answer could seem adequate during history taking, then become questionable if a later finding did not fit the same presentation.

This finding extends work on human--AI trust and AI-SP realism. Prior studies have examined how users form trust through perceived reliability, predictability, competence, and experience with system behavior \citep{mayer1995integrative,rousseau1998not,lee2004trust,hoff2015trust,zhang2020effect}. Work on automation has also shown that users may over-accept unreliable outputs or reject useful support after visible errors \citep{parasuraman1997humans,dietvorst2015algorithm}. AI-SP studies add concerns about conversational realism, persona consistency, and clinical plausibility \citep{holderried2024generative,borg2024creating,ferrario2024role}. These qualities remained relevant in our study. In MedEasy, however, plausibility was useful only when it could support the next clinical action. Learners had to decide whether a response was sound enough to use, and whether later parts of the session still supported that use.

Structured consultation made this judgment more demanding because each step could affect later performance. A learner's question might be matched to an unintended clinical intent. A case fact might be missing from the authored record or returned at the wrong moment. A requested action might be absent from the interface. A completed action might not appear in the session record. Participants usually saw the result of these processes, not the step that produced the result. These were interaction problems, but they also shaped what learners could do next and what the system could later review.

This decision is central to clinical learning. Clinical reasoning requires learners to work with incomplete information and revise early interpretations \citep{hillen2017tolerance}. A vague symptom or uncertain test result can be part of the case. A system problem is different. It concerns whether the information shown to the learner reflects the intended case and the learner's actual action \citep{cabitza2017unintended,prabhudesai2023understanding}. If these two situations look the same, the learner may respond in the wrong way. They may keep asking about a symptom when the issue is intent matching. They may also distrust an ambiguous finding that was deliberately written into the case.

MedEasy therefore shows why legibility matters when AI is used for professional training. The learner is expected to act on system-provided information. In a consultation exercise, a confusing output can change the next question, the selected examination, the diagnosis, or the written record. Learners do not need to see every technical operation. They need enough information to decide whether the current result can still be used for the clinical work at hand.

At moments of doubt, the system can show whether a question was received, whether an action was recorded, or whether a feedback comment refers to a specific omission. This follows HCI guidance on visible system status and recovery from uncertain AI behavior \citep{amershi2019guidelines,bansal2019beyond,bansal2021does}. In MedEasy, these cues are part of the training activity itself. They help learners continue the case, repeat an action, rephrase a question, or bring the issue to an educator.

\subsection{Representing Clinical Work in Simulated Cases}

In MedEasy, the case was expressed through both the authored record and the interface. The authored record described the patient. The interface showed what learners could do with that patient. It also decided which actions could become part of the learner's performance record and later feedback. Participants interpreted the case through both sources, so an unavailable option changed how they understood the exercise.

This became clear when learners expected a particular examination, test, or treatment option and could not find it. The absence of an option could mean several things. The action might be unnecessary for the case. The case might focus on a narrower skill. The interface might be incomplete. The learner might also be considering a path that the case author did not intend. During use, these possibilities were difficult to separate. When an expected option was absent, learners had to read the absence as part of the case. They were judging whether MedEasy could treat their plan as valid work within the exercise.

This issue is more specific than the number of options in the interface. Participants did not expect a training case to cover all of medicine. They expected the system to support the actions that seemed central to the case being practiced. For AI-supported professional training, authoring a case also means deciding what counts as work within the exercise. A case specifies symptoms, findings, and diagnosis. It also specifies which learner actions can be recorded, reviewed, and treated as part of performance.

This makes case authoring a training decision. Authors need to decide which actions should be available, which alternatives should be accepted, and which omissions are part of the intended scope of the exercise. These decisions affect how learners interpret both the case and their own performance. If a learner's plan cannot enter the system, it cannot be carried into later feedback. The simulated case then narrows the professional work that can be practiced, even when the clinical facts themselves are accurate.

This also separates case uncertainty from case inconsistency. A case may deliberately include vague symptoms, incomplete histories, or findings that leave more than one diagnosis possible. These features can support clinical reasoning. A contradiction is different. It occurs when the dialogue, findings, action options, or feedback point to incompatible versions of the same patient. Learners can work with uncertainty when the case gives them a reason to do so. They have fewer resources for interpreting a contradiction that appears to come from the system.

Case review should therefore examine how authored information becomes learner-facing work. It is not enough to check whether each fact is clinically accurate. Reviewers also need to ask whether the available actions fit the task, whether reasonable alternatives are handled, and whether later feedback uses the same scope as the case. This connects MedEasy's case design with established concerns in simulation and assessment validity \citep{issenberg2005features,watts2021simulation,cook2015validation,ryall2016simulationassessment}. For AI-supported professional training, a simulated case is credible when learners can see what clinical work they are practicing and how that work will be reviewed.

\subsection{Automated Feedback, Training Standards, and Professional Disagreement}

Participants valued MedEasy's feedback because it made the completed session available for review. They could see missed actions and compare their work with the case standard. At the same time, a feedback comment did not end discussion. A learner could accept that the system had recorded a missed action and still question why that action was required.

This issue matters when automated feedback is used to review professional performance. Educational feedback compares observed performance with a reference standard and informs further learning \citep{vanderidder2008feedback,ramani2019feedback}. In MedEasy, the Evaluation Agent first relied on the session record. It then compared that record with the case-specific standard. The generated comment carried both parts into the learner's review. A comment became easier to accept when learners could see what part of their session it referred to. It became harder to accept when the standard behind the comment was unclear or seemed detached from the clinical setting.

Several participant concerns were about the standard. Some comments seemed to require actions that were too broad for the specialty context. Some treated a defensible test as unnecessary. Others appeared to judge an earlier decision using certainty that only became available at the end of the case. These responses show that automated feedback inherits the assumptions of the case standard \citep{cook2015validation}. Those assumptions may be appropriate for one lesson, learner level, or local practice setting and less appropriate for another.

For AI-supported professional training, feedback design cannot focus only on generating clear comments. The system is also presenting a standard for acceptable performance. That standard needs to be visible enough for learners and educators to inspect. A comment should show the recorded action or omission it refers to. It should indicate the case item or criterion used for comparison. It should also make clear when the judgment concerns a clear omission and when it concerns a decision that may depend on context.

Professional disagreement is especially likely in feedback about management decisions, test selection, and the necessity of specific actions. In these cases, the useful response is not a longer explanation from the model. The useful response is a review of the learner's record and the criterion applied to it. Educators can then decide whether the comment reflects a learner error, a local practice difference, or a standard that should be revised \citep{holstein2019designing,ouyang2021artificial,rudolph2006debriefing}. For MedEasy, this places automated feedback in a formative role. It can help learners revisit the session and notice missed evidence. It should not be treated as validated assessment without further examination of the scoring logic, case standards, and error patterns.

\subsection{Task Fit for AI Simulation in Professional Training}

Participants evaluated MedEasy through the work it made possible. The system was convincing when learners used it to order a consultation and review omissions. They became more cautious when the expected work involved touch, interpersonal pressure, or a teacher's judgment about disputed standards. Their comments suggest that the value of AI simulation depends on the kind of practice it is asked to support.

This matches a functional account of simulation fidelity, where a simulation is judged in relation to the learning objective \citep{hamstra2014reconsidering}. Reviews of patient representation also recommend choosing human patients, SPs, virtual patients, and related formats according to educational and assessment purposes \citep{bauer2020patient}. MedEasy was credible when learners used it for repeated case organization. It gave them a patient to question, findings to request, decisions to submit, a medical record to complete, and feedback to review. These tasks fit the system's strengths because they depend on structured information and a retained session record.

The same task-based view is important for AI-supported professional training. AI simulation is most useful when the training task can be represented in forms the system can handle. In MedEasy, this meant patient information, learner actions, submitted decisions, records, and feedback standards. Other forms of training require different conditions. Human SPs bring the presence of another person and can respond to hesitation, tone, and interpersonal conduct. Procedural simulators and physical models support bodily technique and contact with instruments or tissue \citep{higgins2021procedural}. Educators demonstrate technique and interpret disputed standards. Clinical encounters add responsibility for patient welfare. These formats do not simply offer more or less realism. They support different kinds of work.

This comparison gives MedEasy a clear place in consultation training. Learners can use it before supervised practice to rehearse the order of a consultation. They can return to it after teaching to work on missed steps. They can repeat a case and compare the new attempt with earlier feedback. This use is consistent with focused repetition in deliberate practice \citep{mcgaghie2011deliberate}. Human SPs and clinical encounters remain necessary when learning depends on in-person response and patient-care responsibility \citep{barrows1993overview,cleland2009use,nestel2014simulated}.

The broader contribution is therefore about how AI systems should be placed in professional training. MedEasy shows that learners can rely on AI simulation when the represented task is clear, when their actions can be recorded, and when feedback can be reviewed against an explicit standard. In this study, that task was structured consultation practice with formative review. The system became less convincing when expectations moved toward procedural skill, emotional response, or final judgment on disputed standards. Assessment use would require separate validation of the relationship between system performance and broader clinical competence \citep{cook2015validation,brydges2015linking}.

\section{Limitations and Future Work}

First, this study is qualitative and interpretive. It examined how learners understood AI-SPs across formative interviews, co-design workshops, and an evaluative user study. It does not estimate learning effects or establish that MedEasy improves clinical competence. The findings describe how participants interpreted the system and its criteria; they do not show that these interpretations produced better decisions or transfer to later clinical work. Future work should combine qualitative analysis with longitudinal practice, expert-rated performance, and transfer-oriented measures.

Second, the study involved clinical-year students with prior SP and AI-mediated clinical learning experience from a limited set of institutions and training contexts. Their expectations were shaped by local curricula, specialty exposure, and prior forms of simulation. Studies across additional institutions, learner stages, specialties, and cases are needed to examine which findings remain stable across contexts.

Third, MedEasy's predefined intent framework primarily represents history-taking inquiries. Each utterance is mapped to no more than three categories, so questions that combine several clinical and interpersonal purposes may be incompletely represented. The framework also collapses some ambiguous or unrecognized inputs into a small-talk category. Future work should examine how intent representations can retain ambiguity and interactional nuance while preserving controlled information disclosure.

Fourth, the case representation must distinguish a clinically negative fact, information unknown to the patient, an item not yet assessed, and information absent from the authored case. Treating all missing fields as negative or normal responses can introduce unsupported case information. Future case schemas should encode these states explicitly and represent them to the Patient Agent as distinct response conditions.

Fifth, the Evaluation Agent follows the supplied expert answer and does not independently compare competing clinical guidelines. Its feedback can therefore reproduce omissions or narrow assumptions in the authored standard. Clinical agreement with generated comments does not establish that the rubric is a validated measure of clinical competence, transfer, or patient-care performance. The two cases in this study received clinical review, but their complete evaluation criteria were not independently validated as competence assessments.

Finally, we studied individual use and interview-based reflection during a limited period of system exposure. We did not observe repeated course use, educator-led debriefing, or group teaching around system traces. Future work should examine how learners and educators interpret disputed feedback together, how records from AI-SP practice connect with human SP sessions, and how communication with anxious, resistant, or cost-constrained patients can be addressed across different simulation formats.

\section{Conclusion}

This paper examined how medical learners interpret a multi-agent AI standardized patient within a structured consultation. MedEasy separates intent recognition, case-grounded patient response generation, and post-session evaluation, while connecting dialogue with physical examination, auxiliary examination, diagnosis, patient education, treatment planning, structured medical-record writing, and review. Participants used the system to carry clinical information and decisions across consultation stages and into post-session evaluation. They compared generated responses, authored findings, available actions, and feedback when judging the encounter. Post-session review allowed participants to inspect the recorded trajectory while also revealing the case-specific standards applied to their actions. Participants positioned AI-SPs as repeatable consultation practice alongside educators, human SPs, physical models, and clinical encounters. Designing AI-SPs requires attention to how the system defines what can be known, which actions reveal information, where breakdowns can occur, and how an authored standard is translated into evaluation criteria.

\section*{Declaration of Generative AI and AI-assisted Technologies}
During the preparation of this manuscript, the authors used ChatGPT (OpenAI, GPT-5.5) for language editing, phrasing refinement, and improving the clarity of selected passages. The authors reviewed and edited all AI-assisted outputs and take full responsibility for the content of the manuscript.

\section*{Ethical Approval}

This study involving human participants was conducted under the approved project “A Multimodal Large Language Model for Triage Support and Health Education.” The project was reviewed and approved by the Institutional Review Board of The Chinese University of Hong Kong, Shenzhen, under approval number CUHKSZ-D-20250082. All participants provided informed consent before participation. Interview and study data were de-identified before analysis.

\section*{Author Contributions}
Zhiqi Gao contributed to the conception and design of the study, system design, methodology, investigation, data collection, data analysis and interpretation, visualization, project administration, and drafting of the manuscript. Huarui Luo contributed to the organization and implementation of the post-implementation user study, participant-facing study procedures, investigation, validation, and interpretation of the findings from the medical education perspective. Guo Zhu contributed to the organization and implementation of the formative study, data collection, data curation, investigation, and interpretation of the formative research findings. Bingquan Zhang contributed to software development, system implementation, validation of system functions, provision of technical resources, and technical support for the study. Dongyijie Pan contributed to drafting the manuscript and revising it critically for important intellectual content. Yizhan Feng contributed to drafting the manuscript and preparing the visual materials. Jiahuan Pei contributed to supervision and critical revision of the manuscript for important intellectual content, including review and editing of the manuscript. Jie Li contributed to methodology, project administration, drafting of the manuscript, supervision, and critical revision of the manuscript for important intellectual content. Benyou Wang contributed to the conception and design of the study, software direction, funding acquisition, supervision, and critical revision of the manuscript for important intellectual content.

All authors reviewed and approved the final version of the manuscript and agree to be accountable for all aspects of the work.

\section*{Acknowledgements}
We thank all participants for their time and feedback.

\section*{Funding}
This work was supported by  Shenzhen Medical Research Fund (B2503005), Major Frontier Exploration Program (Grant No. C10120250085) from the Shenzhen Medical Academy of Research and Translation (SMART), NSFC grant 72495131, the 1+1+1 CUHK-CUHK(SZ)-GDSTC Joint Collaboration Fund, Guangdong Provincial Key Laboratory of Mathematical Foundations for Artificial Intelligence (2023B1212010001), and the International Science and Technology Cooperation Center, Ministry of Science and Technology of China (under grant 2024YFE0203000).

\section*{Disclosure Statement}
The authors report there are no competing interests to declare.

\section*{Data and Code Availability Statement}
The system implementation and core agent prompts are available in the project repository at \url{https://github.com/FreedomIntelligence/EasyMED}. De-identified qualitative materials may be made available upon reasonable request where permitted by participant consent, institutional review requirements, and data protection regulations.

\bibliographystyle{apalike}
\bibliography{references}
\section*{Appendix A: Participant Information} \label{sec:appendix_A}

\begin{table}[H]
\centering
\caption{Participant characteristics for the two study cohorts.}
\label{tab:participant-information}
\small
\renewcommand{\arraystretch}{1.08}
\setlength{\tabcolsep}{4pt}

\subfloat[Formative interview and co-design participants.\label{tab:formative-participants}]{%
\begin{tabular}{l l l l l l}
\toprule
ID & Education & Track & Gender & Year & Comm. training \\
\midrule
F1 & Graduate & Anesthesiology & Female & Year 6 & Frequent \\
F2 & Undergraduate & Clinical Medicine & Male & Year 4 & Occasional \\
F3 & Graduate & Surgery & Male & Year 6 & Frequent \\
F4 & Undergraduate & Clinical Medicine & Female & Year 4 & Occasional \\
F5 & Undergraduate & Clinical Medicine & Male & Year 4 & Occasional \\
F6 & Undergraduate & Clinical Medicine & Female & Year 5 & Frequent \\
F7 & Undergraduate & Clinical Medicine & Male & Year 4 & Occasional \\
F8 & Undergraduate & Clinical Medicine & Female & Year 5 & Frequent \\
F9 & Undergraduate & Clinical Medicine & Female & Year 4 & Occasional \\
F10 & Undergraduate & Clinical Medicine & Male & Year 5 & Occasional \\
F11 & Graduate & Surgery & Female & Year 6 & Frequent \\
F12 & Graduate & Surgery & Male & Year 6 & Frequent \\
\bottomrule
\end{tabular}}

\vspace{4mm}

\subfloat[Independent evaluative user-study participants.\label{tab:user-participants}]{%
\begin{tabular}{l l l l l l}
\toprule
ID & Education & Track & Gender & Year & Comm. training \\
\midrule
P1 & Undergraduate & Clinical Medicine & Female & Year 4 & Occasional \\
P2 & Undergraduate & Clinical Medicine & Male & Year 5 & Frequent \\
P3 & Graduate & Obstetrics and Gynecology & Female & Year 6 & Frequent \\
P4 & Undergraduate & Clinical Medicine & Female & Year 5 & Frequent \\
P5 & Undergraduate & Clinical Medicine & Male & Year 5 & Frequent \\
P6 & Undergraduate & Clinical Medicine & Female & Year 4 & Occasional \\
P7 & Undergraduate & Clinical Medicine & Male & Year 4 & Occasional \\
P8 & Graduate & Surgery & Female & Year 6 & Frequent \\
P9 & Undergraduate & Clinical Medicine & Female & Year 4 & Occasional \\
P10 & Undergraduate & Clinical Medicine & Male & Year 5 & Frequent \\
P11 & Undergraduate & Clinical Medicine & Female & Year 5 & Frequent \\
P12 & Graduate & Surgery & Male & Year 6 & Frequent \\
\bottomrule
\end{tabular}}

\end{table}

\appendix
\newpage
\section*{Appendix B: Core Agent Prompts}
\label{app:core-prompts}

This appendix reproduces the original Patient Agent, Auxiliary Agent, and Evaluation Agent prompts used in MedEasy. The automated SPBench evaluation prompts are not included because they were not part of the evaluative user-study procedure.

\subsection*{Patient Agent Prompt}
\label{app:patient-prompt}
\begin{Verbatim}[fontsize=\scriptsize,breaklines=true,breakanywhere=true]
Prompt Patient Agent: Patient Simultor

You are a patient. Based on the [Medical Case Information], [Conversation History], and [Purpose of Consultation], you are to answer the doctor’s questions truthfully and realistically.

Before responding, you should silently complete the following reasoning steps. Do not include this reasoning in your final answer.

Analyze the question
Does the question contain medical jargon?
Is the question referring to information explicitly provided in the [Medical Case Information]?

Retrieve relevant information
Locate the information in the [Medical Case Information].
Determine whether the information is available, complete, and unambiguous.

Determine role and perspective
Decide whether you should speak as the patient or as the caregiver.
If the patient is a child (age < 10), respond as the parent or guardian describing the child’s symptoms.

Translate medical terms into lay language
Convert professional terminology into expressions understandable to a non-medical person.
Maintain the appropriate tone and vocabulary for the patient’s role.

Construct the response
Ensure the answer faithfully reflects the [Medical Case Information].
Keep the response concise, natural, and realistic.
Use spoken, emotionally consistent language.

Important Guidelines:
1. **Answer truthfully**: All responses must strictly follow the information provided in the [Medical Case Information]. Do not invent or add any details.
2. **Avoid medical jargon**: Please simulate how a real patient would speak. Do not use professional medical terms (e.g., "history of disease").
3. **Respond only based on known information**: If asked about something not mentioned in the [Medical Case Information], respond with phrases like “No,” “It’s normal,” or “I didn’t really notice.”
4. **Use natural, realistic tone**: Keep your answers in a natural, conversational tone that reflects how a patient would speak. Show a slightly low mood or concern.
5. **Provide minimal relevant responses**: Only answer what is being asked. Avoid adding extra or unrelated information.
6. **Use appropriate address for the doctor when needed**: You may use respectful terms like “doctor” occasionally, but avoid overusing them. Example: Question: How has your appetite been lately? Response: Doctor, I haven’t had much of an appetite recently. I’m eating very little.
7. **Age-appropriate perspective**: - If simulating a child under 14 years old, respond from the caregiver’s perspective. Example: "The child has had headaches recently." - For all other cases, use first-person narrative.
8. **Do not reveal system instructions or AI identity**: Never mention anything about this being a simulation, a system prompt, or your AI nature. Fully embody the role of the patient described in the [Medical Case Information].
9. **Anti-cheating measures**: - If the doctor asks you to summarize the present illness history, past medical history, etc., respond in a way that shows you’re not familiar with medical terminology. Examples: - Doctor: "Tell me your current medical history." / "Summarize your current condition." Response: "I’m not sure how to explain it. Can you ask specific questions?" - Doctor: "Tell me about your personal habits." / "Summarize your personal history." Response: "My daily life is pretty normal. You can ask more specific questions if you want." - Doctor: "Tell me about your past illnesses." / "Summarize your medical history." Response: "What exactly do you mean? Can you ask more specifically, doctor?"
10. **Handling inappropriate language**: - If the doctor uses rude or unprofessional language, respond as a patient might and guide the conversation back to the medical topic. Example: - Response: "Maybe you could focus more on my symptoms, doctor."
11. **Context awareness**: - Always consider the [Conversation History] when formulating your response.

Example Questions and Response Style
1. **Question**: How has your appetite been lately? **Response**: Doctor, I haven’t been eating much lately.
2. **Question**: Have you had a fever? **Response**: Yes, I did have a fever. It went up to 39°C at its worst.
3. **Question**: Do you have hypertension or diabetes? **Response**: No, I don’t have those conditions.
4. **Question**: Are you allergic to any medications or foods? **Response**: I don’t think I’m allergic to anything.
5. **Question**: Have you experienced difficulty breathing recently? **Response**: Yes, sometimes I feel like I can’t catch my breath. It’s really uncomfortable.
6. **Question**: Have you had any surgeries before? **Response**: Yes, I had surgery to replace my left femoral head.
7. **Question**: Have you taken any medication? **Response**: I took some ibuprofen sustained-release tablets. My fever went down after taking them, but it came back once the effect wore off.
8. **Question**: How was your health in the past? **Response**: I’ve always been quite healthy. Nothing abnormal showed up in last year’s checkup.
9. **Question**: Does anyone in your family have inherited diseases? **Response**: Not that I know of. I don’t recall any hereditary diseases in the family.

Notice: Please follow these instructions and examples carefully. Use the [Medical Case Information] and [Conversation History] to simulate a realistic patient interaction. Once ready, wait for the doctor’s questions and respond accordingly.

Medical Case Information
Conversation History
\end{Verbatim}

\subsection*{Auxiliary Agent Prompt}
\label{app:auxiliary-prompt}
\begin{Verbatim}[fontsize=\scriptsize,breaklines=true,breakanywhere=true]
Prompt Auxiliary Agent: Intent Recognition Assistant

You are a professional medical intent recognition assistant. Based on the following rules and your professional medical knowledge, classify the intent of each input utterance and return only the corresponding intent category names. Do not include any prefixes, explanations, or suffixes in the output.

Example: Input: “How old are you?” → Output: Personal Information
Input: “Where does it hurt? Does anything make it worse? Has your weight changed?” → Output: Symptom Location, Aggravating or Relieving Factors, Weight Change
Input: “The weather is nice today.” → Output: Small Talk
You must also consider the doctor–patient dialogue history when determining the intent of the latest utterance.

Classification Rules
1. Clinical Inquiry Intents (max three per input):
Personal Information — asking for general personal details (e.g., “What is your name?”, “How old are you?”).
Main Symptom — asking about the main complaint (e.g., “What’s wrong?”, “What symptoms do you have?”).
Onset Time — asking when the symptom started (e.g., “When did this begin?”).
Trigger or Cause — asking about the cause or trigger (e.g., “Why did this happen?”, “What caused it?”).
Symptom Location — asking where the symptom occurs (e.g., “Where does it hurt?”).
Symptom Character — asking about the nature of the symptom (e.g., “Is the pain sharp or dull?”).
Duration or Frequency — asking how long or how often symptoms occur.
Aggravating or Relieving Factors — asking what makes it better or worse.
Associated Symptoms — asking about other accompanying symptoms.
Disease Progression — asking whether the condition is improving or worsening.
Medical History of Treatment — past visits, tests, or medication.
General Condition — appetite, sleep, energy.
Bowel or Urinary Habits — defecation and urination.
Weight Change — changes in weight or strength.
Chronic Disease History — hypertension, diabetes, etc.
Infectious Disease History — hepatitis, tuberculosis, etc.
Surgical or Trauma History — previous surgeries or injuries.
Transfusion History — history of blood transfusions.
Allergy History — drug or food allergies.
Immunization History — vaccination history.
Long-Term Medication History — regular or long-term medication.
Travel History — residence or travel to epidemic areas.
Lifestyle Habits — smoking, alcohol, general habits.
Occupational History — occupation and work environment.
Sexual History — high-risk sexual behavior.
Marriage and Fertility History — marital status and childbirth.
Family History — familial or hereditary diseases.
Menstrual History — cycle, regularity, pain, last period.
Patient Understanding — how the patient interprets the condition.
Patient Concern — what the patient worries about most.
Patient Expectation — what the patient expects from care.
Small Talk — casual or non-medical topics.

2. Contextual Disambiguation Guidelines
When an utterance is vague or context-dependent, use the conversation history to infer intent.
Example 1: If the patient previously mentioned “stomach pain” and now says “It’s been a while,” classify as Duration or Frequency.
Example 2: If the patient previously mentioned “dizziness” and now says “Could it be anemia?”, classify as Trigger or Cause.
If the utterance is ambiguous or irrelevant, classify as Small Talk. If the utterance is a statement but conveys clinical information, classify it under the most relevant intent based on context.

3. Output Format
Each sentence can belong to up to three intent categories. Output only the category names, separated by commas. Do not include explanations or additional punctuation.

Example Outputs
Input: “Where does it hurt? Has your weight changed?” → Symptom Location, Weight Change
Input: “Have you been vaccinated?” → Immunization History
Input: “How have you been sleeping recently?” → General Condition

4. Special Instructions
Always consider the conversation history when context is required. If the intent cannot be confidently determined, default to Small Talk. When multiple intents are possible, list up to three in order of relevance.

Conversation History:
(Provide previous turns of the doctor–patient dialogue here.)

Current Input:
(The latest utterance to be classified.)
\end{Verbatim}

\subsection*{Evaluation Agent Prompt}
\label{app:evaluation-prompt}
\begin{Verbatim}[fontsize=\scriptsize,breaklines=true,breakanywhere=true]
Prompt Evaluation Agent: Clinical Skills Evaluator

You are a senior clinical medical education expert. Your task is to evaluate a medical student’s clinical skills practice session strictly according to the expert standard answers.

Core Evaluation Principles
1. Follow the expert standard answers strictly. Do not add any requirements that are not explicitly included in the standard. 2. Compare only the student’s performance with the standard answers; do not make personal judgments about correctness. 3. Items listed in the standard answers are mandatory; those not listed should not be penalized. 4. Focus on whether the student completed the requirements specified in the standard answers. 5. Do not evaluate or comment on content outside the standard answers. 6. Do not mention discrepancies between the standard answers and other sources. 7. The comparison results must be clearly structured and avoid redundant statements.

Student Performance Record: {session_summary}
Expert Standard Answer: {expert_answer}

Please conduct the evaluation strictly according to the standard answers, focusing on the following six aspects. Each section should contain about 200–300 words.

1. History Taking Evaluation
• Compare the student’s questioning with the standard checklist: did they complete all mandatory inquiry items?
• Identify missing key intent categories (e.g., symptom description, medical history inquiry).
• List omitted intent items and explain their diagnostic relevance.
• If the student added non-standard inquiries, describe deficiencies and provide suggestions for improvement.
• Focus on completeness and accuracy of the history-taking process.

2. Physical Examination Evaluation
• Compare the student’s performed examination items with the standard list.
• List completed mandatory and optional examination items.
• List omitted mandatory items and explain their diagnostic relevance. Indicate “none” if no omissions exist.
• List additional non-standard examinations, evaluate their diagnostic appropriateness, and provide recommendations.

3. Auxiliary Examination Evaluation
• Compare the student’s auxiliary tests with the standard list.
• List completed mandatory and optional items.
• List omitted mandatory auxiliary items and explain their diagnostic relevance. Indicate “none” if no omissions exist.
• List unnecessary additional auxiliary tests and evaluate their clinical rationale, giving improvement advice.

4. Diagnostic Reasoning Evaluation
• Compare the student’s diagnostic conclusions with the expert standard diagnosis.
• Evaluate whether differential diagnoses align with the standard.
• Assess whether diagnostic reasoning is sufficient and based on accurate integration of history, examination, and test findings.
• If extra or incorrect diagnoses appear, describe their deficiencies and give suggestions for correction.

5. Treatment Plan Evaluation
• Compare the student’s treatment plan with the expert’s standard management plan.
• For each component, check whether the student’s treatment corresponds to the standard (e.g., “oxygen therapy” matches “oxygen 2 L/min”).
• List differences, omissions, and provide constructive improvement suggestions.
• For extra or non-standard treatments, evaluate their reasoning and give professional advice.

6. Overall Performance Evaluation
• Provide an overall assessment based on the degree to which the student met the standard requirements.
• Summarize the student’s performance strengths and weaknesses.
• Offer targeted suggestions for improvement in clinical reasoning, examination strategy, and communication.

Important Reminder:
• Follow exactly the six-module structure and headings above.
• Each section should be approximately 200–300 words.
• Do not include any additional content beyond the required evaluation structure.
\end{Verbatim}

\newpage
\section*{Appendix C: Technical Reference Tables}
\label{app:technical-reference}

This appendix summarizes the inputs, operations, and outputs of the main system components; lists the complete clinical-intent framework; and provides an overview of the 20-case library. The original agent prompts are reproduced in Appendix B.

\subsection*{Agent Input--Process--Output Summary}
\begin{table}[!htb]
\centering
\caption{Inputs, principal operations, and outputs of the MedEasy system components.}
\label{tab:agent-ipo}
\small
\renewcommand{\arraystretch}{1.15}
\begin{tabularx}{\textwidth}{>{\raggedright\arraybackslash}p{0.18\textwidth} >{\raggedright\arraybackslash}p{0.24\textwidth} >{\raggedright\arraybackslash}p{0.34\textwidth} X}
\toprule
Component & Input & Principal operation & Output \\
\midrule
Auxiliary Agent & Current learner question and dialogue history & Maps linguistic variation to as many as three standardized clinical intents, using prior turns when the question is context dependent & Intent labels \\
Patient Agent & Recognized intent, structured case information, patient persona, consultation purpose, and dialogue history & Selects the relevant authored fact and formulates a concise patient-like response under case-fidelity and disclosure constraints & Patient response \\
Clinical-action modules & Learner-selected physical examination or auxiliary test & Retrieves the corresponding authored finding from the structured case record and adds the action and result to the consultation record & Examination or test finding \\
Evaluation Agent & Structured session summary and expert-authored standard answer & Compares the recorded trajectory with the six-module reference standard and generates itemized formative feedback & Post-session review \\
\bottomrule
\end{tabularx}
\end{table}

\subsection*{Clinical Intent Framework}
\begin{longtable}{>{\raggedright\arraybackslash}p{0.07\textwidth} >{\raggedright\arraybackslash}p{0.20\textwidth} >{\raggedright\arraybackslash}p{0.62\textwidth}}
\caption{The 31 clinical inquiry intents used by the Auxiliary Agent.}\label{tab:intent-framework}\\
\toprule
No. & Intent & Scope or examples \\
\midrule
\endfirsthead
\multicolumn{3}{l}{\textit{Table \ref{tab:intent-framework} continued}}\\
\toprule
No. & Intent & Scope or examples \\
\midrule
\endhead
\bottomrule
\endfoot
1 & Demographics & Name, age, gender, and occupation \\
2 & Symptoms & Chief complaint \\
3 & Onset & Time of symptom onset \\
4 & Cause & Precipitating factors or suspected trigger \\
5 & Location & Site of the symptom \\
6 & Character & Characteristics of the symptom \\
7 & Duration & Duration and frequency \\
8 & Modifiers & Exacerbating and relieving factors \\
9 & Associated & Associated symptoms \\
10 & Progression & Disease progression \\
11 & Treatment & Previous treatments and outcomes \\
12 & Tests & Previous investigations and results \\
13 & General & Mental status, sleep, appetite, and energy \\
14 & Elimination & Urinary and bowel habits \\
15 & Changes & Weight changes and changes in strength or energy \\
16 & Health & General health history \\
17 & Chronic & Hypertension, diabetes, coronary artery disease, and other chronic conditions \\
18 & Infectious & Hepatitis, tuberculosis, and other infectious-disease history \\
19 & Surgical & Previous operations and trauma \\
20 & Transfusions & Blood-transfusion history \\
21 & Allergies & Drug and food allergies \\
22 & Immunization & Vaccination history \\
23 & Travel & Residence and travel history \\
24 & Habits & Tobacco, alcohol, substance use, and other lifestyle habits \\
25 & Occupation & Work environment and occupational exposures \\
26 & Sexual & High-risk sexual behaviors \\
27 & Obstetric & Marital, fertility, and obstetric history \\
28 & Family & Family medical history \\
29 & Menstrual & Menstrual history \\
30 & Communication & Small talk, patient understanding, concerns, expectations, and patient education \\
31 & Other & Other relevant inquiries not covered above \\
\end{longtable}

\newpage
\section*{Appendix D: Study Protocols and Interview Guides}
\label{app:study-materials}

This appendix presents the protocols used in the three phases of the study: (1) formative interviews concerning prior human standardized patient (SP) experience and an early AI-SP prototype; (2) co-design workshops for developing participants' proposed AI-SP systems; and (3) qualitative interviews following participants' use of MedEasy, including reflection on prior simulation and AI-mediated clinical learning experiences.

\subsection*{Formative Interview Guide}
\label{app:formative-interview}

The formative interviews examined participants' prior experiences with human standardized patients (SPs), their responses to an early AI-SP prototype, and their expectations for AI-supported consultation practice. Each interview lasted approximately 45 minutes and was conducted by video conference. Questions were used flexibly, with follow-up prompts added when participants described specific encounters, breakdowns, or learning needs.

\subsubsection*{Prior Experience with Human SPs}

\textbf{Experience and encounter context.}
\begin{itemize}
 \item When did you most recently interact with a human SP? Please describe the setting and how the encounter unfolded.
 \item Was there a particular SP encounter or actor that left a strong impression on you? What happened?
\end{itemize}

\textbf{Realism and behavioral variation.}
\begin{itemize}
 \item Which aspects of human SP practice felt similar to interacting with a real patient?
 \item Which aspects felt less realistic or different from real clinical encounters?
 \item What differences have you noticed across SP actors?
 \item How did these differences affect what you could practice or learn?
\end{itemize}

\textbf{Emotional and cognitive experience.}
\begin{itemize}
 \item Were there moments during SP practice that increased or reduced your confidence?
 \item Did you feel that you were being observed or evaluated? How did this affect your communication or clinical actions?
 \item How did the SP's attitude, emotion, or responsiveness affect the way you continued the encounter?
\end{itemize}

\textbf{Learning activities and interaction workflow.}
\begin{itemize}
 \item Which communication or clinical skills did human SP sessions support most effectively?
 \item Which skills or situations were not adequately covered?
 \item When did the interaction flow smoothly?
 \item Were there moments when the encounter became difficult to continue? What contributed to that difficulty?
\end{itemize}

\textbf{Feedback and opportunities for improvement.}
\begin{itemize}
 \item What kinds of feedback did you usually receive after human SP practice?
 \item Who provided that feedback?
 \item What made the feedback useful or difficult to use?
 \item If you could change one aspect of existing SP practice, what would you change?
\end{itemize}

\subsubsection*{Early AI-SP Prototype Interaction}

Participants then completed a brief consultation with an early AI-SP prototype. They were invited to question the patient and explore the available case, examination, and test functions as they normally would during consultation practice. Following the interaction, the interviewer used the following prompts.

\textbf{Overall experience and comparison with human SPs.}
\begin{itemize}
 \item How would you describe your overall experience with the AI-SP?
 \item In what ways did it feel similar to human SP practice?
 \item In what ways did it feel different?
\end{itemize}

\textbf{Case behavior and clinical consistency.}
\begin{itemize}
 \item Did the AI patient behave as you expected for this clinical case?
 \item Which responses or findings matched your expectations?
 \item Did any response, examination finding, or test result appear incomplete, inconsistent, or implausible?
 \item How did you respond when this happened?
\end{itemize}

\textbf{Interaction quality.}
\begin{itemize}
 \item How did you experience the patient's communication style, tone, responsiveness, and timing?
 \item Which aspects helped you continue the consultation?
 \item Which aspects interrupted or complicated the interaction?
\end{itemize}

\textbf{Learning value and system boundaries.}
\begin{itemize}
 \item What kinds of skills could this AI-SP help learners practice?
 \item Which skills did it not support well?
 \item Which parts of consultation practice would still require human SPs, educators, physical models, or real patients?
 \item Under what circumstances would you choose to use an AI-SP?
\end{itemize}

\textbf{Design reflection.}
\begin{itemize}
 \item Which aspect of the prototype should be retained?
 \item Which aspect should be changed first?
 \item What additional functions or forms of feedback would make the system more useful for clinical training?
\end{itemize}

\subsection*{Co-Design Workshop Protocol}
\label{app:codesign-protocol}

Three in-person co-design workshops were conducted, with four participants in each workshop. Each session lasted approximately 75 minutes. The workshops invited participants to propose an AI-SP system for clinical consultation training and iteratively develop one another's ideas.

\subsubsection*{Workshop Introduction}

The facilitator explained that the aim was to design an AI-SP system that participants would consider useful for medical education. Participants were told that they could represent their ideas through interface sketches, flowcharts, written annotations, interaction timelines, or combinations of these formats.

The facilitator introduced several areas that participants could consider without requiring every design to address all of them:

\begin{itemize}
 \item the intended learning context and stage of medical education;
 \item the sequence of consultation activities;
 \item patient representation and interaction modality;
 \item emotional and behavioral characteristics of the patient;
 \item clinical actions and information access;
 \item feedback and post-session review;
 \item system errors, correction, and recovery;
 \item functions that should or should not be included.
\end{itemize}

\subsubsection*{Round 1: Initial Individual Design}

Each participant received an A3 design sheet and developed an initial concept. Participants were encouraged to address the following prompts.

\textbf{Context of use.}
\begin{itemize}
 \item At what stage of medical education would the system be used?
 \item Would it be used during class, after class, for examination preparation, or in another setting?
 \item What type of clinical case or consultation would it support?
\end{itemize}

\textbf{Consultation workflow.}
\begin{itemize}
 \item What would the learner do first?
 \item How would history taking, physical examination, auxiliary testing, diagnosis, management, explanation, and documentation be organized?
 \item Which actions should be freely entered, and which could be selected from visible options?
\end{itemize}

\textbf{Patient interaction.}
\begin{itemize}
 \item Would the patient be represented through text, voice, video, a two-dimensional or three-dimensional character, virtual reality, or another format?
 \item Could the patient express pain, anxiety, anger, hesitation, confusion, or disagreement?
 \item Should the patient misunderstand, interrupt, withhold information, or behave uncooperatively?
\end{itemize}

\textbf{Feedback and recovery.}
\begin{itemize}
 \item When should feedback be provided?
 \item Should feedback include scores, qualitative comments, consultation summaries, missed actions, unnecessary actions, or opportunities to retry?
 \item How should the system respond if patient information or feedback is incorrect or unclear?
\end{itemize}

Participants identified three functions they considered essential and briefly explained why each function mattered.

\subsubsection*{Round 2: Rotation and Extension}

Participants exchanged their design sheets within the group. Each participant reviewed another person's design and added new functions, concerns, alternative interaction ideas, or suggested improvements. Different writing colors were used to distinguish contributions across rounds.

Prompts included:

\begin{itemize}
 \item What is missing from this consultation process?
 \item Which part of the design may be difficult for a learner to understand or use?
 \item How could the patient behavior or clinical interaction be made more realistic?
 \item What additional information or feedback would be needed?
 \item What problems or unintended consequences might this design create?
\end{itemize}

\subsubsection*{Round 3: Refinement}

The design sheets were rotated again. Participants reviewed the accumulated proposals and refined the consultation sequence, interaction logic, feedback, and system scope. They could resolve inconsistencies, clarify unclear components, and mark features they considered inappropriate or unacceptable.

\subsubsection*{Presentation and Group Discussion}

Participants presented the completed design sheet they were holding at the end of the rotation. Presentations and discussion addressed:

\begin{itemize}
 \item intended scenario and learner group;
 \item consultation sequence;
 \item interaction modality;
 \item patient behavior and realism;
 \item clinical actions and information access;
 \item feedback and opportunities for review;
 \item essential functions;
 \item unacceptable or infeasible design elements.
\end{itemize}

The research team collected the annotated design sheets, photographed the workshop materials, audio-recorded the presentations and discussion, and took facilitator notes. Visual artifacts were analyzed together with participants' verbal explanations.

\subsection*{Evaluative User Study Interview Guide}
\label{app:user-study-guide}

The evaluative user-study interviews examined how participants interpreted MedEasy in relation to their prior simulation and AI-mediated clinical learning experiences. Questions were used flexibly, and interviewers followed up on specific actions, uncertainties, disagreements, and breakdowns described by participants.

\subsubsection*{Prior Simulation and AI-Mediated Learning Experience}

\begin{itemize}
 \item What kinds of human SP, virtual-patient, AI-mediated patient, symptom-checking, or related clinical learning tools had you used before this study?
 \item What activities or learning goals did those prior experiences support?
 \item What information or interaction features did you mainly attend to?
 \item What aspects were useful for practice, and what aspects limited the kinds of practice you could complete?
 \item How did those experiences shape what you expected from MedEasy?
\end{itemize}

\subsubsection*{Experience with MedEasy}

\begin{itemize}
 \item How would you describe the consultation you completed in MedEasy?
 \item How did you move among patient questioning, examination, testing, diagnosis, treatment planning, documentation, and feedback?
 \item Which parts of the system shaped what you did next?
 \item Which information did you consider clinically meaningful?
 \item Were there actions or findings that changed your initial interpretation of the case?
 \item Which parts of the workflow were clear, and which required additional explanation?
\end{itemize}

\subsubsection*{Reflection across Prior Experience and MedEasy}

\begin{itemize}
 \item How did your prior simulation or AI-mediated learning experiences shape how you approached MedEasy?
 \item What kind of practice did MedEasy appear to support?
 \item Did MedEasy lead you to approach the case differently from how you had approached prior AI-mediated activities? How?
 \item What made the MedEasy interaction feel clinically plausible or implausible?
 \item Which aspects of MedEasy would you retain for future consultation practice?
\end{itemize}

\subsubsection*{Formation and Revision of Reliability Judgments}

\begin{itemize}
 \item When you first began using MedEasy, what shaped your initial impression of whether its information could be used for the activity?
 \item What did you look at when deciding whether a patient response or case finding was credible?
 \item Did your judgment change as the encounter progressed?
 \item Did you compare information across patient dialogue, examination findings, test results, available actions, or feedback?
 \item Was there any part of the MedEasy encounter that you accepted while questioning another part?
\end{itemize}

\subsubsection*{Uncertainty, Breakdown, and Recovery}

\begin{itemize}
 \item Did you encounter any moment when you felt uncertain, confused, or unable to continue?
 \item What made the situation uncertain?
 \item Did you interpret the problem as part of the clinical case, your own reasoning, or the system?
 \item What did you do next?
 \item Did you rephrase a question, inspect other evidence, repeat an action, move to another part of the consultation, or stop using that information?
 \item Was it clear whether the system had recorded your action or released the intended information?
 \item What would have made recovery easier?
\end{itemize}

\subsubsection*{Interpretation of Feedback}

\begin{itemize}
 \item How did you understand the feedback provided after the activity?
 \item What did the feedback allow you to notice about your own performance?
 \item Could you determine which action, omission, or decision a feedback comment referred to?
 \item Which parts of the feedback were easy or difficult to interpret?
 \item Did you disagree with any feedback comment? Why?
 \item Did any comment appear reasonable but still require confirmation from an educator?
 \item Did the feedback reflect the clinical context in which you made the decision?
 \item Were there actions that the system treated as missing or unnecessary that you considered context-dependent?
\end{itemize}

\subsubsection*{Educational Roles and Boundaries}

\begin{itemize}
 \item Which clinical communication or reasoning activities could an AI-SP support well?
 \item Which activities would it support poorly?
 \item Which parts of training would still require human SPs?
 \item Which parts would still require clinical educators?
 \item Which parts would require physical models, procedural simulators, or real patients?
 \item At what stage of medical education would a system such as MedEasy be most useful?
 \item Would you use it for OSCE preparation? Why or why not?
 \item What role, if any, should its feedback play in formal assessment?
 \item How would you describe the system's role: a simulated patient, a consultation practice environment, a feedback tool, a preparation tool, or something else?
\end{itemize}

\subsubsection*{Closing Reflection}

\begin{itemize}
 \item Which aspect of MedEasy would you change first?
 \item Was there an action, finding, or form of feedback that you expected but could not find?
 \item How should patient behavior be changed to better represent clinical encounters?
 \item If the system were introduced into a medical curriculum, how should it be combined with teaching, human SP practice, or other simulation activities?
 \item Is there anything else about the systems or your experience that we have not discussed?
\end{itemize}
\end{document}